\documentclass[twocolumn]{aastex63}

\usepackage{placeins}
\usepackage{amsmath}

\newcommand*{\teff}{T_{\mathrm{eff}}}
\newcommand*{\mines}{\texttt{MINESweeper}}
\newcommand*{\ang}{\mbox{\AA}}

\begin{document}

\title{The Galactic Distribution of Phosphorus: A Survey of 163 Disk and Halo Stars\footnote{Based on observations obtained with the Hobby-Eberly Telescope, which is a joint project of the University of Texas at Austin, the Pennsylvania State University, Ludwig-Maximilians-Universit\"{a}t M\"{u}nchen, and Georg-August-Universit\"{a}t G\"{o}ttingen.}}

\author[0000-0002-0475-3662]{Zachary G. Maas} 
\affil{McDonald Observatory, The University of Texas at Austin, 2515 Speedway Boulevard, Austin, TX 78712, USA}
\affil{Department of Astronomy, The University of Texas at Austin, 2515 Speedway Boulevard, Austin, TX 78712, USA}

\author[0000-0002-1423-2174]{Keith Hawkins} 
\affil{Department of Astronomy, The University of Texas at Austin, 2515 Speedway Boulevard, Austin, TX 78712, USA}

\author{Natalie R. Hinkel} 
\affil{Southwest Research Institute, 6220 Culebra Road, San Antonio, TX 78238, USA}

\author{Phillip Cargile}
\affil{Center for Astrophysics $\mathrm{|}$ Harvard $\&$ Smithsonian, 60 Garden Street, Cambridge, MA 02138, USA}

\author{Steven Janowiecki} 
\affil{University of Texas, Hobby-Eberly Telescope, McDonald Observatory, TX 79734, USA}

\author{Tyler Nelson} 
\affil{Department of Astronomy, The University of Texas at Austin, 2515 Speedway Boulevard, Austin, TX 78712, USA}

\email{zgmaas@utexas.edu}

\begin{abstract}

Phosphorus (P) is a critical element for life on Earth yet the cosmic production sites of P are relatively uncertain. To understand how P has evolved in the solar neighborhood, we measured abundances for 163 FGK stars over a range of --1.09 $<$ [Fe/H] $<$ 0.47 using observations from the Habitable-zone Planet Finder (HPF) instrument on the Hobby-Eberly Telescope (HET). Atmospheric parameters were calculated by fitting a combination of astrometry, photometry, and Fe I line equivalent widths. Phosphorus abundances were measured by matching synthetic spectra to a P I feature at 10529.52 $\ang$. Our [P/Fe] ratios show that chemical evolution models generally under-predict P over the observed metallicity range. Additionally, we find that the [P/Fe] differs by $\sim$ 0.1 dex between thin disk and thick disk stars that were identified with kinematics. The P abundances were compared with $\alpha$-elements, iron-peak, odd-Z, and s-process elements and we found that P in the disk most strongly resembles the evolution of the $\alpha$-elements. We also find molar P/C and N/C ratios for our sample match the scatter seen from other abundance studies. Finally, we measure a [P/Fe] = 0.09 $\pm$ 0.1 ratio in one low-$\alpha$ halo star and probable Gaia-Sausage-Enceladus (GSE) member, an abundance ratio $\sim$ 0.3 - 0.5 dex lower than the other Milky Way disk and halo stars at similar metallicities. Overall, we find that P is likely most significantly produced by massive stars in core collapse supernovae (CCSNe) based on the largest P abundance survey to-date.
\end{abstract}


\section{Introduction}
\label{sec::intro}

The only stable isotope of P ($^{31}$P) is thought to be created in multiple different nucleosynthesis channels. Empirical tests of these different production processes are limited by the few known stellar P abundances. A low galactic abundance of P (relative to the even-Z $\alpha$-elements) and no detectable optical transitions in FGK-type stars have made large-scale abundance studies challenging. The first P chemical evolution study, totaling 20 stars, used P I lines spanning 10500 - 10600 $\ang$ in the infrared (IR) Y-band \citep{caffau11}. Recently, additional studies have  used the IR Y-band lines \citep{caffau16,maas17,maas19,caffau19,sneden21,Sadakane22}, UV P I lines at $\sim$ 2134 $\mbox{\AA}$ \citep{roederer14,jacobson14,roederer16,spite17}, and lines in the IR H-band from $\sim$ 15700 - 16500 $\ang$ \citep{hawkins16,afsar18,topcu19,jonsson18,jonsson20,masseron20a}. 

These surveys combine for a total $\sim$ 190 P abundances, with an average of 21 stars per study. One notable exception is the APOGEE survey which reports P abundances for tens of thousands of stars. However the IR H-band lines are blended with other molecular features and impacted by telluric contamination, making P the most uncertain element with reported abundances in the APOGEE catalog and should be used with caution \citep{hawkins16,jonsson20}. Nonetheless, the APOGEE survey has led to the discovery of P-rich stars \citep{masseron20a}. At higher resolutions, the IR H-band lines may be used for more precise P abundance measurements (e.g., with the IGRINS instrument, \citealt{afsar18,topcu19}).

The initial $\sim190$ stellar P abundances have placed some constraints on galactic chemical evolution models. The Y-band lines can be used to study stars with [Fe/H] $\gtrsim$ --1.3 and the UV lines are best detected in metal-poor stars. The combination of these studies have demonstrated that hypernovae best describe the evolution of P at early epochs \citep{jacobson14} and phosphorus abundances for stars spanning --1 $<$ [Fe/H] $<$ 0.5 do not match chemical evolution models tuned for the solar neighborhood, that include yields from CCSN, Type Ia SN, and asymptotic giant branch (AGB) stars (e.g. \citealt{cescutti12,maas19}). The current best fitting model uses arbitrarily enhanced massive star yields from \citet{kobayashi06} to match existing measurements (model 8 in \citealt{cescutti12}). The limited number of P abundances have demonstrated theoretical models are either missing additional P-production process, P yields must be increased, or possibly both. 

While P is thought to be primarily created by neutron capture on Si during hydrostatic burning and released in core collapse supernovae CCSNe \citep{woosley95}, and an enhanced P abundance as been observed in the supernova remnant Cas A \citep{koo13}, other processes such as massive rotating stars \citep{prantzos18} and C-O shell mergers within stars \citep{ritter18} are promising methods to enhance P yields. However when these processes are utilized in chemical evolution models, they do not yet match the P abundances observed in the galaxy. Analysis of APOGEE abundances has suggested P may be made in Type Ia SN \citep{weinberg19} or P may be made via proton capture in massive stars \citep{caffau11}. Additionally, P-rich stars have also challenged theoretical models; abundance patterns of these stars includes enhanced Mg, Si, Al, and s-process elements but are not consistent with other chemically peculiar stars \citep[e.g. CH stars,][]{masseron20a,masseron20b}. The unusual abundances suggest P-enriched stars may be from a new site of the s-process \citep{masseron20b}. 

Beyond chemical evolution, P is an important element for life: phosphorus-bearing molecules make up the backbone of DNA/RNA \citep{schlesinger13} and are key in energy transport between cells \citep{nelson17}. However, of the key elements for life (the CHNOPS elements), phosphorus is the most poorly studied in stars \citep{jacobson14,hinkel20}. P is especially important as this element may be strongly partitioned in the cores of planets \citep{stewart07}, so an under-abundance of P relative to the solar value may impact the available surface P on exoplanets \citep{hinkel20}. While the chemical composition of exoplanet surfaces cannot be directly measured, in certain cases host star abundances can be used as a proxy \citep{thiabaud15} and the first step to understanding the availability of P on other planets is to measure P in stars \citep{hinkel20}. 

Understanding the chemical evolution of P in the galaxy, the nucleosynthesis sites of P, and the connection between P in stars and planets relies on stellar abundance measurements. To this end, we have measured stellar abundances in 163 stars using the HPF instrument on the HET, creating the largest survey of precise P abundances to-date. Our observations sample selection is described in section \ref{sec::sample}. Atmospheric parameter and abundance derivation methodology is discussed in section \ref{sec::atmo_params}. The key results that include the galactic chemical evolution of P, likely important P nucleosynthesis processes, [P/Fe] comparisons between thin and thick disk stellar populations, and P abundances translated to molar fractions are discussed in in section \ref{sec::discussion}. The primary conclusions are listed in section \ref{sec::conclusions}.

\section{Sample Selection and Observations}
\label{sec::sample}

\begin{figure}[tp!]
	\centering 
 	\includegraphics[trim=0cm 0cm 0cm 0cm, scale=.4]{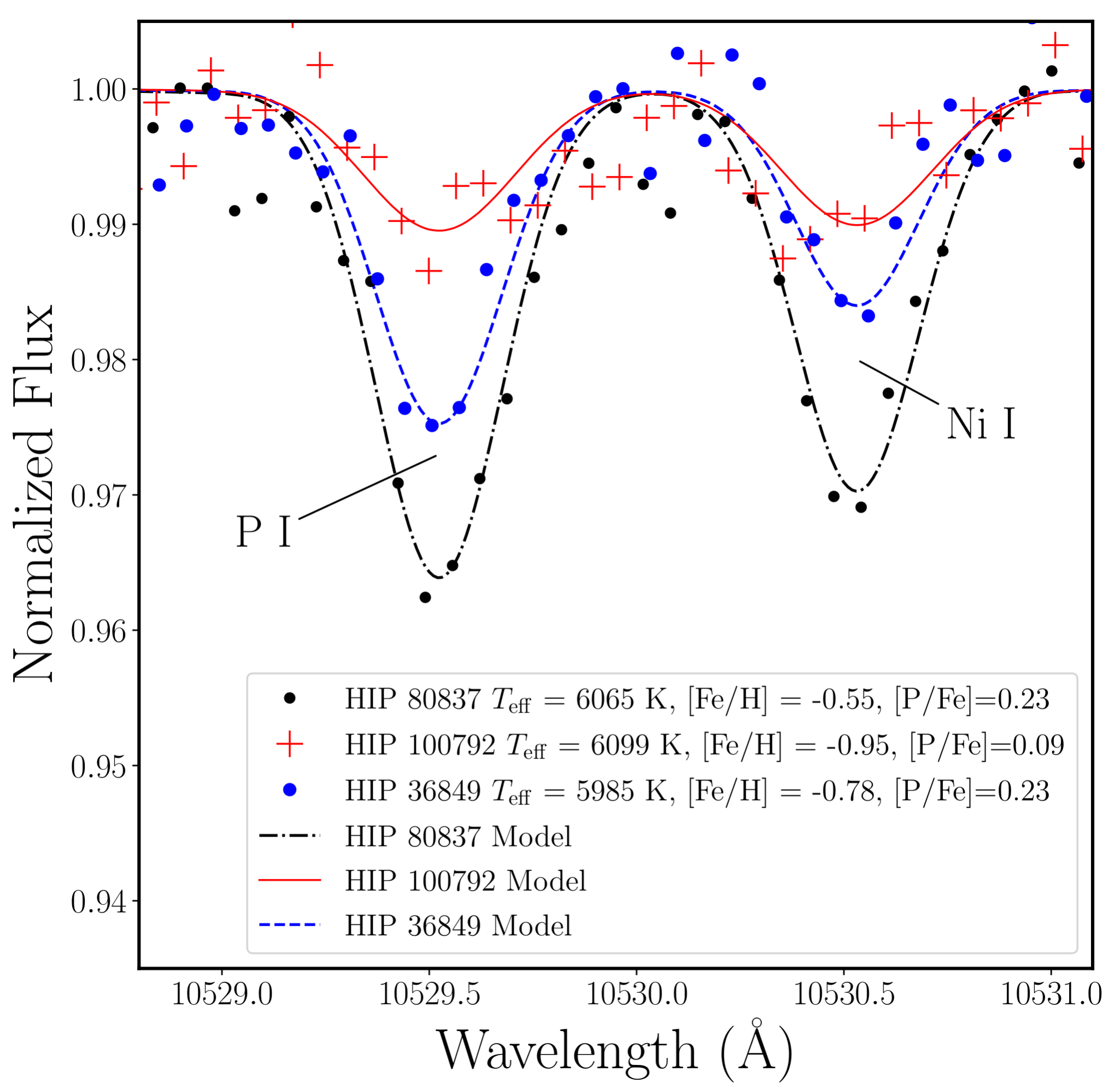}
	\caption{Three observations of relatively metal-poor stars with similar $\teff$ and log(g) from HPF on the HET are plotted with three synthetic spectra. The observed spectrum of HIP 80837 is represented as black circles, HIP 100792 is shown as red crosses, and HIP 36849 is shown as blue circles. Model spectra for these stars are displayed as a black dot-dashed line, a red solid line, and a blue dashed line respectively. \label{fig:ir_spec} }
	\end{figure}

IR P I features have weak line strengths relative to other absorption lines (as shown by three representative HPF spectra in Fig. \ref{fig:ir_spec}) and are difficult to detect for stars with --1.3 $\lesssim$ [Fe/H] \citep{maas19}. The stellar sample was thus chosen to ensure the Y-band P I features were observable. Our target stars were selected from \citet{bensby14} and APOGEE DR16 \citep{jonsson20}. Stars with J $\lesssim$ 8 mag were chosen from both surveys to minimize needed exposure times to achieve signal-to-noise ratios (SNRs) over 100, which are required to detect the P I feature (e.g. \citealt{maas19}). The P I features in the Y-band have high excitation potentials of $\sim$ 7 eV, and become difficult to detect in cool stars (e.g. Arcturus, \citealt{maas17}). We only selected stars with $\teff$ $>$ 4500 K to ensure the P I feature was detectable and consistent with the $\teff$ range used in previous studies \citep{maas17,maas19}. 

Observations were performed using the HPF spectograph \citep{mahadevan12,mahadevan14} on the HET. HPF is fiber-fed and observes an approximate wavelength range of 8000 - 13000 $\ang$ with a resolution (R) $\sim$ 50,000. Target stars were observed between Aug. 2020 - Dec. 2020 with metal poor and/or thick disk targets, identified with [Fe/H] from \citet{bensby14,jonsson20} and initial population memberships from \citet{bensby14}, prioritized for observation. A total of 163 stars were observed with the HPF with photometry, observation dates, SNRs, and parallaxes given for each star in Table \ref{table::obslog}. 

Data reduction was performed using the \texttt{Goldilocks} automatic pipeline\footnote{\url{hydra.as.utexas.edu/?a=help&h=121}}. The pipeline reduction steps follow \citet{ninan18} and can be summarized in the following steps: first the bias level was subtracted using reference pixels on the edges of the detector. Next, non-linearity corrections, a scattered light subtraction, and a flat field correction were performed. Science and calibration images were extracted using a master trace image. The final data product included a sky fiber spectrum, target spectrum, errors, and wavelength solution for each star. We do not perform a telluric correction to maximize the final SNR. Previous analysis has shown spectra from the \texttt{Goldilocks} pipeline lead to accurate and precise stellar abundances (e.g. \citealt{sneden21}). 

The strongest P I feature with no telluric contamination in the HPF spectrum is at 10529.52 $\ang$ and is the primary P I line used for abundance measurements in this study. Other lines are not available due to gaps in the wavelength coverage or blends with telluric lines. Additionally, there are two sky emission lines relatively close to the P I features. Each object spectrum was visually inspected and the sky emission was corrected using a sky fiber spectrum when the lines were nearly blended with the 10529.52 $\ang$ P I feature. Only $\sim$ 10 stars required sky subtraction to remove contamination from or near the P I feature line profile. The spectra were normalized using a spline function with a sigma-clipping algorithm. The SNR of each data was found by estimating the variance from featureless regions of the continuum near the P I feature. 

\begin{splitdeluxetable*}{ c c c c c c c  c c c c c c c  c B c c c c c c c c c c c c c}
\tablewidth{0pt} 
\tabletypesize{\scriptsize}
\tablecaption{Sample Parallaxes, and Observation Summary \label{table::obslog}} 
 \tablehead{\colhead{Name} &\colhead{Date Obs.} & \colhead{RA.} & \colhead{DEC.} & \colhead{Obs}  &\colhead{G} & \colhead{$\sigma$G}& \colhead{BP} & \colhead{$\sigma$BP} & \colhead{RP} & \colhead{$\sigma$RP}& \colhead{VT} & \colhead{$\sigma$VT}& \colhead{BT} & \colhead{$\sigma$BT}& \colhead{J} & \colhead{$\sigma$J}& \colhead{H} & \colhead{$\sigma$H} & \colhead{K$_{s}$}  & \colhead{$\sigma$K$_{s}$}& \colhead{W1}  & \colhead{$\sigma$W}& \colhead{W2} & \colhead{$\sigma$W2} & \colhead{$\pi$}  & \colhead{$\sigma \pi$ }    \\
 \colhead{} & \colhead{(UT Date)} & \colhead{deg.}& \colhead{(deg.)} & \colhead{(SNR)} &  \colhead{(mag)} &  \colhead{(mag)} &  \colhead{(mag)} &  \colhead{(mag)} &  \colhead{(mag)} &  \colhead{(mag)} &  \colhead{(mag)} &  \colhead{(mag)} &  \colhead{(mag)} &  \colhead{(mag)} &   &  \colhead{(mag)} &  \colhead{(mag)} &  \colhead{(mag)} &  \colhead{(mag)} &  \colhead{(mag)} &  \colhead{(mag)} &  \colhead{(mag)} &  \colhead{(mag)}  &  \colhead{(mag)} &\colhead{(mas)}&  \colhead{(mas)}}
\startdata
HIP101346 & 2020-08-07 & 308.09979 & 6.51752 & 283.0 & 8.401 & 0.003 & 8.675 & 0.003 & 7.950 & 0.004 & 8.56 & 0.01 & 9.15 & 0.01 & 7.44 & 0.02 & 7.17 & 0.02 & 7.10 & 0.02 & 7.08 & 0.03 & 7.11 & 0.02 & 10.04 & 0.022 \\
HIP102610 & 2020-08-21 & 311.90774 & 12.98616 & 209.0 & 7.296 & 0.003 & 7.590 & 0.003 & 6.825 & 0.004 & 7.50 & 0.01 & 8.13 & 0.01 & 6.28 & 0.02 & 5.97 & 0.02 & 5.91 & 0.02 & 5.87 & 0.05 & 5.82 & 0.03 & 24.94 & 0.018 \\
HIP102838 & 2020-08-07 & 312.51246 & 7.85000 & 192.0 & 8.393 & 0.003 & 8.713 & 0.003 & 7.898 & 0.004 & 8.61 & 0.01 & 9.36 & 0.01 & 7.35 & 0.02 & 7.04 & 0.05 & 6.95 & 0.03 & 6.90 & 0.03 & 6.96 & 0.02 & 16.74 & 0.020 \\
HIP103682 & 2020-08-27 & 315.14089 & -4.72995 & 158.0 & 6.064 & 0.003 & 6.369 & 0.003 & 5.590 & 0.004 & 6.26 & 0.01 & 6.96 & 0.01 & 5.15 & 0.03 & 4.86 & 0.02 & 4.77 & 0.02 & 4.72 & 0.08 & 4.46 & 0.04 & 37.00 & 0.029 \\
HIP103692 & 2020-08-02 & 315.18085 & 17.44797 & 116.0 & 7.895 & 0.003 & 8.230 & 0.003 & 7.388 & 0.004 & 8.13 & 0.01 & 8.91 & 0.01 & 6.82 & 0.03 & 6.53 & 0.05 & 6.48 & 0.02 & 6.44 & 0.04 & 6.46 & 0.02 & 17.55 & 0.021 \\
HIP104659 & 2020-08-10 & 317.99626 & 17.72986 & 233.0 & 7.231 & 0.003 & 7.501 & 0.003 & 6.781 & 0.004 & 7.43 & 0.01 & 7.95 & 0.01 & 6.25 & 0.02 & 5.99 & 0.03 & 5.94 & 0.02 & 5.91 & 0.05 & 5.84 & 0.02 & 29.88 & 0.020 \\
HIP104672 & 2020-08-22 & 318.03117 & 10.63799 & 129.0 & 7.948 & 0.003 & 8.222 & 0.003 & 7.502 & 0.004 & 8.13 & 0.01 & 8.74 & 0.01 & 7.03 & 0.02 & 6.78 & 0.03 & 6.69 & 0.02 & 6.68 & 0.04 & 6.70 & 0.02 & 12.74 & 0.023 \\
\enddata
\tablecomments{Sources of photemetry and parallax: J, H, and K$_{s}$ Source 2MASS \citep{twomass_ref,skrutskie06}, VT and BT from \citet{esa97,hog00}, W1 and W2 photometry from WISE \citep{wright10,wise_ref}, and G, BP, RP, parallax ($\pi$) and parallax uncertainty from Gaia eDR3 \citep{gaia16,gaia21}.}
\tablecomments{(This table is available in its entirety in machine-readable form.)}
\end{splitdeluxetable*}

\section{Atmospheric Parameters}
\label{sec::atmo_params}
\subsection{Methodology Overview}
We required an atmospheric parameter derivation methodology that was applicable to both cool K giants and F dwarfs. Typical iron line excitation analysis (e.g. Fe I vs. Fe II) to derive $\teff$ and log(g) were difficult to adapt for the Y-band spectra as there are few detectable Fe II lines, especially for the cool K giants, within the observed spectral range 8000 - 13000 $\ang$ \citep{kondo19,sneden21}. Additionally, while there has been significant progress in Y-band abundance methodologies (e.g. \citealt{caffau19,kondo19,matsunaga20,fukue21,sneden21}), atomic line data has not been experimentally validated at the same level as for optical wavelengths. We therefore combine information from multiple surveys to determine the atmospheric parameters using photometry, parallaxes, and finally equivalent width measurements of Fe I lines in our spectral range. 

The effective temperature, surface gravity, and their uncertainties were calculated using the $\mines$ spectrophotometric modeling tool\footnote{github.com/pacargile/MINESweeper} \citep{cargile20}. The software is capable of simultaneously fitting an observed stellar spectrum and isochrones to predict various stellar atmospheric parameters using a dynamic nested sampler \citep{speagle20}. \mines\ has been validated using multiple star clusters \citep{cargile20} and has been used to determine parameters for the H3 survey \citep{conroy19}. We only utilized the photometric fitting capabilities due to the added uncertainties found when deriving atmospheric parameters broadly from Y-band absorption features. $\mines$ makes use of MESA Isochrones and Stellar Tracks (MIST) \citep{choi16}, which were created using the Modules for Stellar Astrophysics (MESA) software suite \citep{paxton11}. $\mines$ also utilizes \texttt{ThePayne} \citep{ting19} to interpolate between different stellar models. Finally, $\mines$ uses ATLAS model atmospheres \citep{kurucz79,kurucz13} which are consistent with MARCS models for the metal-rich FGK stars considered in this study \citep{gustafsson08}.

\begin{deluxetable}{ c c c c c}
\tablewidth{0pt} 
\tabletypesize{\footnotesize}
\tablecaption{Linelist \label{table::linelist}} 
 \tablehead{\colhead{Wavelength} & \colhead{Line ID} &  \colhead{$\chi$} &  \colhead{Log(gf)} &  \colhead{Source} \\ \colhead{($\ang$)} & \colhead{} & \colhead{(eV)} & \colhead{} & \colhead{}}
\startdata
8207.742   & Fe I & 4.446 & --0.960	&  (1) \\
8339.402 & Fe I & 4.473 & --1.100    &	 (1) \\
8360.794 & Fe I &  4.473 & --1.100   &	 (1) \\
8471.743 & Fe I & 4.956 & --0.910    & (1) \\
8514.072 & Fe I & 2.198 & --2.227    &	 (1) \\
8571.804 & Fe I & 5.010 &  --1.110    &	(1) \\
8582.257 & Fe I & 2.990 & --2.133    &	(1) \\
8598.829 & Fe I & 4.387 & --1.182    &	(1) \\
8610.610 & Fe I & 4.435 & --1.690    &	(1) \\
8674.746 & Fe I & 2.832 & --1.783    &	 (1) \\
8699.454 & Fe I & 4.956 & --0.370    & (1) \\
8824.220 & Fe I & 	2.198 &  --1.540    &(1) \\
8876.024 & Fe I & 5.0200 &  --1.100    &	(1) \\
10216.313 & Fe I & 4.7330 & --0.062   &	 (2) \\
10395.796 & Fe I & 2.1760 &  --3.390 &	(2) \\
10529.524  & P I &  6.949  & 0.24 & (3) \\
\enddata
\tablecomments{Sources: (1) \citet{heiter21}: (2) \citet{placco21}:  (3) \citet{berzinsh97}}
\end{deluxetable}

\subsection{Input Data and Parameters For MINESweeper}
\label{subsec::minesweeper_input}
We use sets of photometric data spanning from the optical to the infrared to derive the stellar atmospheric parameters. Optical photometry from Gaia eDR3: G, BP, and RP, \citep{gaia16,gaia21}, the Tycho survey: VT and BP \citep{esa97,hog00}, 2MASS: J, H, K$_{s}$ \citep{twomass_ref,skrutskie06}, and WISE: W1 and W2 \citep{wright10,wise_ref} were used with $\mines$. Following \citet{cargile20}, an error floor of 0.01 mag was applied to photometric measurements to ensure the uncertainties were not underestimated. A Gaussian prior was set with the mean and standard deviation equal to the Gaia eDR3 parallaxes and uncertainties respectively \citep{gaia21} for our sample stars. 

The photometric data and parallaxes are given in Table \ref{table::obslog}. We performed an initial $\mines$ run with the input parameters and a [Fe/H] prior set from --1 $<$ [Fe/H] $<$ 0.2. A flat prior on dust extinction was also set at 0 $<$ A$_{v}$ $<$ 0.2 mag for all but one star, HD 25532, which is predicted to have a high extinction of A$_{v}$ $\sim$ 0.35 mag from the python package \texttt{dustmaps} \citep{green19}. Increasing the flat prior boundary to 0.5 mag brought atmospheric parameters for that one star in agreement with literature measurements. We also corrected the Gaia eDR3 parallax zero points using the prescription from \citet{lindegren21}. The average offset and standard deviation was --0.033 $\pm$ 0.003 mas. We applied the average offset for stars that had two-point astrometric solutions and we found the parallax offset was not a significant source of uncertainty for our sample of stars.

Next, we measured [Fe/H] ratios using Fe I lines equivalent widths from lines successfully used in \citet{sneden21}. Equivalent widths were measured by fitting Gaussian profiles to absorption features with the spectrum analysis code \texttt{iSpec} \citep{blanco-cauresma14,blanco-cauresma19}. All lines were crossmatched with a telluric catalog within \texttt{iSpec} and features with significantly blended with telluric lines removed. The equivalent width measurements were tested using Fe I lines from a twilight spectrum with a SNR of $\sim$ 300.  We measured an A(Fe I)$_{\odot}$ = 7.46 $\pm$ 0.05 with a microturbulence ($\xi$) of 0.67 km s$^{-1}$ using a MARCS model atmosphere built to match the Sun ($\teff$ 5777 K and log(g) = 4.44 dex). The Fe I abundance is consistent with previous solar Fe abundance measurements \citep{asplund09} and our Fe I linelist is given in Table \ref{table::linelist}. 

The Fe I abundance for each star was calculated using the local thermodynamic equilibrium (LTE) radiative transfer code MOOG \citep{sneden73} with 1-D MARCS atmospheric models \citep{gustafsson08}. We used plane-parallel model atmospheres for stars with log(g) $>$ 3.5 dex and spherical models for stars with log(g) $<$ 3.5 dex. The initial MARCS model atmospheric parameters were set to the $\mines$ output $\teff$, log(g), [Fe/H], and an assumed microturbulence of 1.0 km s$^{-1}$. We then calculated the Fe I abundances with MOOG and proceeded to perform a grid search over potential microturbulence values to find which best removes any differences in Fe I abundance for both strong and weak lines. However, for giant stars, additional lines with lower excitation energies become saturated at lower effective temperatures and are not used in calculating the final [Fe/H] abundance. We therefore do not calculate the microturbulence using Fe I line equivalent widths for giant stars since the number of stronger Fe I lines were limited. We instead adopt the methodology of APOGEE DR13 \citep{holtzman18} and use an empirical relationship between the microturbulence and  log(g), given in Eq. \ref{eq::logg}, for stars with log(g) $<$ 3.5 dex.

\begin{equation}
\label{eq::logg}
\begin{split}
\mathrm{log}(\xi)=0.225 - 0.0228(log(g)) + 0.0297(log(g))^{2}
\\  - 0.0113(log(g))^{3}
\end{split}
\end{equation}

We then re-run $\mines$ with the new Fe I abundance, and an Fe I prior equal to the [Fe/H] uncertainty. After the second iteration, we removed any lines that give an A(Fe) $>$ 0.2 dex from the average values to remove outliers that may have been impacted by unflagged telluric lines or were saturated. We iterated this process multiple times until the parameters no longer changed with each iteration. Often the parameters did not significantly change after two iterations. The final atmospheric parameters and uncertainties are given in Table \ref{table::params_abuns}.

\begin{figure*}[tp!]
	\centering 
 	\includegraphics[trim=0cm 0cm 0cm 0cm, scale=.33]{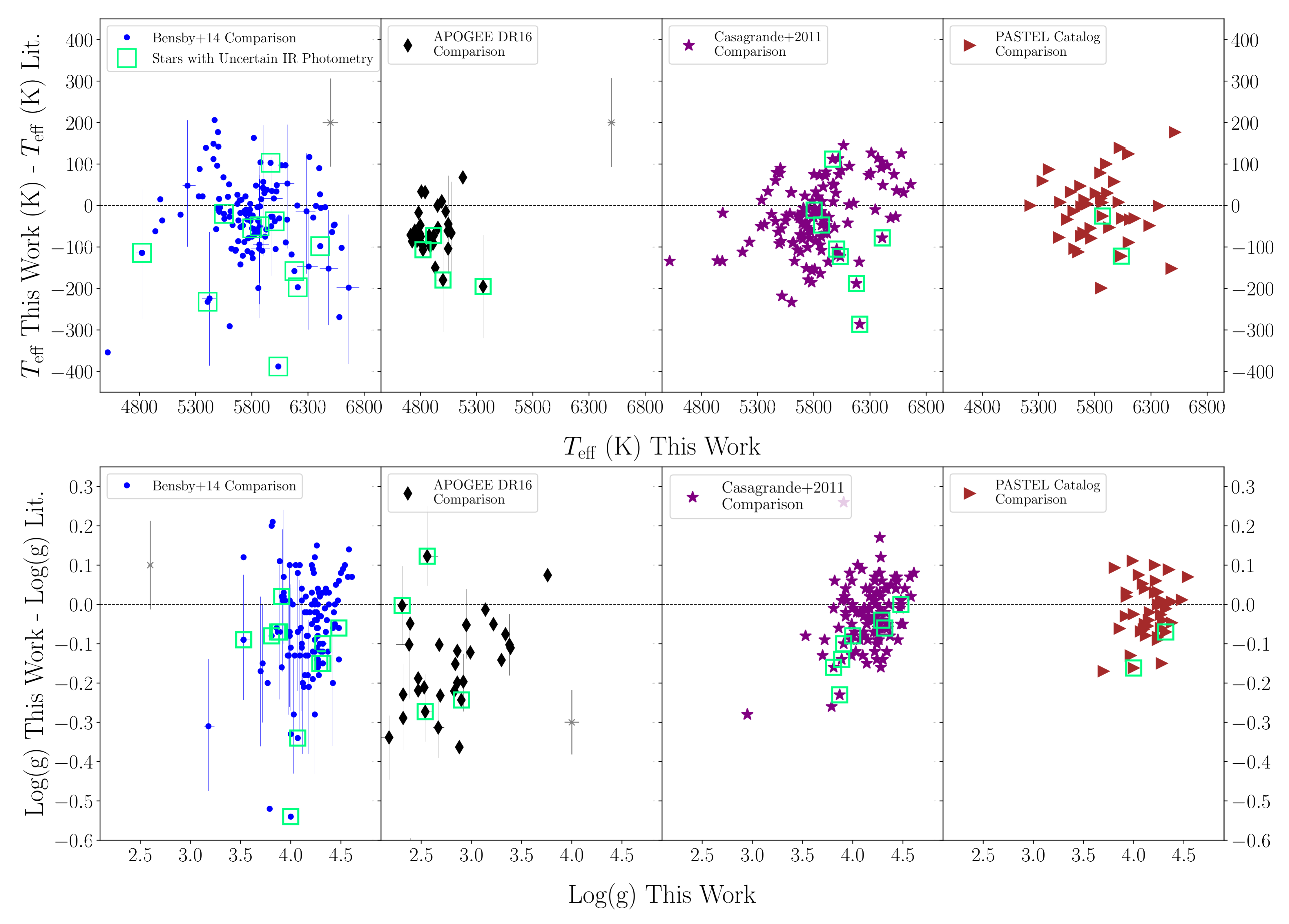}
	\caption{A comparison between this work an \citet{bensby14} is shown as blue circles, with APOGEE DR16 \citep{jonsson20} in the next panel as black diamonds, comparisons to \citet{casagrande11} are magenta stars, and to the PASTEL catalog as brown triangles \citep{soubiran16}. The top panels show the differences in derived effective temperature and the bottom set of panels show differences in log(g). Error bars are only shown for stars with errors above 1$\sigma$ from the average uncertainty on the parameter difference for the \citet{bensby14} and APOGEE DR16 comparisons. We also include representative error bars as gray `x' symbols in the APOGEE DR16 and \citet{bensby14}. Light green boxes highlight stars with highly uncertain ($>$ 0.2 mag) 2MASS photometry for two or more filters. } \label{fig:teff_logg} 
\end{figure*}

The uncertainties on $\teff$ and log(g) are derived from the output posterior distributions from $\mines$. The mean A(Fe I) abundance from the lines used in Table \ref{table::linelist} was taken to be the final iron abundance. The uncertainties on the [Fe/H] ratio is the standard deviation from the set of line equivalent width measurements for star in our sample. We estimated the uncertainty on the microturbulence for stars with log(g) $>$ 3.5 dex by finding which microturbulence value corresponds to the one $\sigma$ uncertainty on the slope of the fit to the logarithm of the line strength and Fe I line abundance. The average uncertainty is 0.25 $\pm$ 0.1 km s$^{-1}$ from this methodology. We also propagated the errors in Eq. \ref{eq::logg} for a log(g) = 3 $\pm$ 0.1 dex. We find an uncertainty of 0.14 km s$^{-1}$, although this error propagation does not include uncertainty on the polynomial fit coefficients. We therefore adopt an error of 0.25 km s$^{-1}$ for all stars in our sample. The iron abundance for each star was then normalized with solar abundances from \citet{asplund09}, A(Fe)$_{\odot}$ = 7.5. The default solar abundances used for MOOG v. 2019 are \citet{asplund09} abundances and therefore we normalized our Fe abundances with this scale. The final atmospheric parameters are listed in Table \ref{table::params_abuns}.

\begin{figure*}[tp!]
	\centering 
 	\includegraphics[trim=0cm 0cm 0cm 0cm, scale=.37]{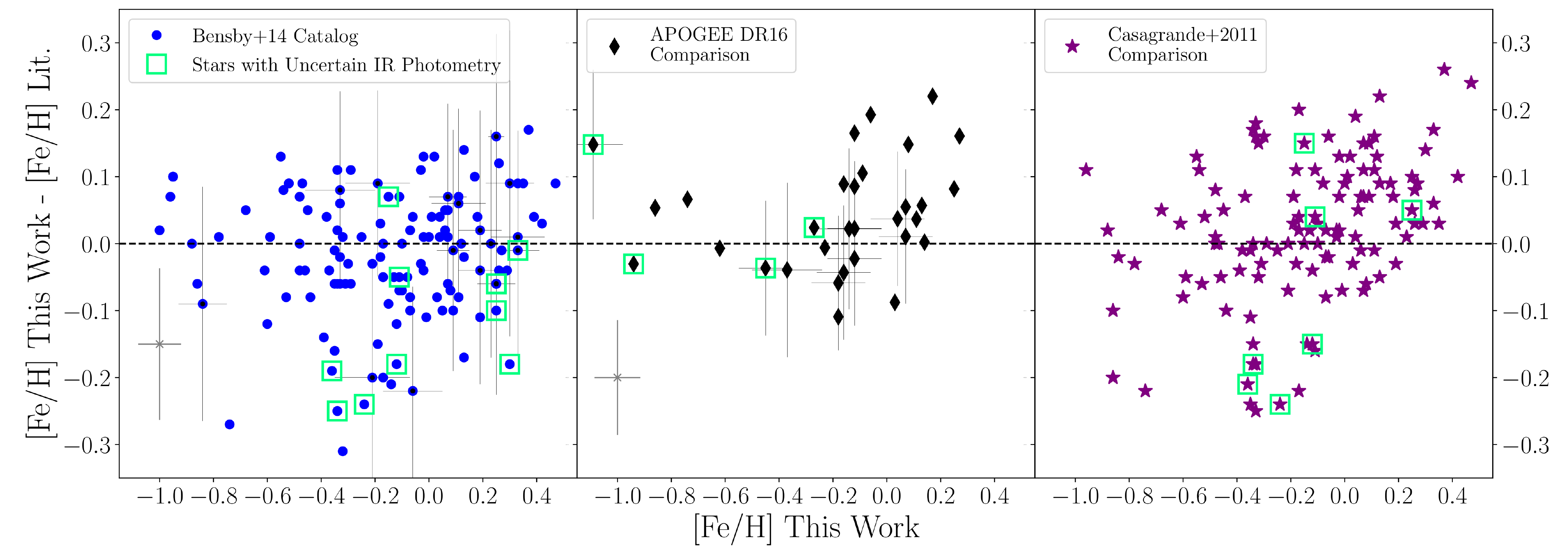}
	\caption{Comparison of literature [Fe/H] abundance measurements to results from this paper and the symbols are the same as Fig. \ref{fig:teff_logg} . \label{fig::feh_comp} }
	\end{figure*}

\subsection{Atmospheric Parameters Comparisons}
\label{subsec::atmo_compare}
We tested the accuracy and precision of our atmospheric parameter methodology by comparing our results to previous literature studies: \citet{bensby14} and APOGEE DR16 \citep{jonsson20}, the primary sources for our sample selection. Additionally, a significant fraction of our stars have been analyzed in other literature studies
and so we also make comparisons to the PASTEL catalog \citep{soubiran16} and a re-analysis of Geneva-Copenhagen targets \citep{casagrande11}. For the PASTEL catalog, effective temperatures and log(g) values used for comparison were the average for all the entries from different literature sources for each star. The comparison of the atmospheric parameters is shown in Fig. \ref{fig:teff_logg}. 

We found our calculated atmospheric parameters are consistent with literature samples for the FGK sample from \citet{bensby14}, as quantified in Table \ref{table::atmo_param_comparisons}. The $\teff$ systematic differences are $\lesssim$ --30 K or less and with standard deviations are similar to the measurement uncertainties for each study. For example, stars from this work had an average $\sigma$ $\teff$ = 62 K and the average $\sigma$ $\teff$ = 72 K from \citet{bensby14} for stars in common. These $\teff$ uncertainties added in quadrature is 95 K, which is similar to the differences seen between both studies in Table \ref{table::atmo_param_comparisons}. Some stars in our sample saturate in one or more of the 2MASS filters which may be the cause of some of the systematic differences observed. We compare stars with only stars with all `AAA' 2MASS quality flags, indicating that J, H, and K$_{s}$ were all reliable measurements and not saturated on the detector. The stars with poor photometry are highlighted in Fig. \ref{fig:teff_logg} and removal of these stars only slightly improves the differences between the different sets of parameters found in the literature, for example the $< \teff >$ offset improves by 10 K for the stars in common with \citet{bensby14} (shown in Table \ref{table::atmo_param_comparisons}). We do find log(g) differences much larger than our reported error and assume a conservative $\sigma$log(g) = 0.1 dex when determining the errors of our [P/Fe] ratios from the uncertainty of the atmospheric parameters (in section \ref{subsection::phos_abun_method}).

\begin{deluxetable*}{ c c c c c c c c c c c}
\tablewidth{0pt} 
\tabletypesize{\footnotesize}
\tablecaption{Atmospheric Parameter and Abundances \label{table::params_abuns}} 
 \tablehead{\colhead{Name} & \colhead{$\teff$} &  \colhead{$\sigma \teff$ High} &\colhead{$\sigma \teff$ Low} &\colhead{log(g)} &  \colhead{$\sigma$ log(g)} & \colhead{[Fe/H] }  & \colhead{$\sigma$ [Fe/H] }   & \colhead{$\xi$}  & \colhead{[P/Fe]} & \colhead{$\sigma$ [P/Fe]}  \\ \colhead{} & \colhead{(K)} & \colhead{(K)} & \colhead{(K)} & \colhead{(dex)}& \colhead{(dex)}& \colhead{(dex)}& \colhead{(dex)}& \colhead{(km s${^{-1}}$)}& \colhead{(dex)} & \colhead{(dex)}}
\startdata
HIP101346 & 6012 & 94& 76& 3.99 & 0.02 & --0.45 & 0.08 & 1.09 & 0.09 & 0.05 \\
HIP102610 & 5902 & 60& 89& 4.27 & 0.02 & --0.11 & 0.08 & 0.89 & --0.01 & 0.05 \\
HIP102838 & 5611 & 71& 34& 4.27 & 0.02 & --0.29 & 0.06 & 0.82 & 0.42 & 0.06 \\
HIP10306 & 6409 & 92& 99& 3.89 & 0.02 & --0.34 & 0.06 & 1.85 & 0.09 & 0.03 \\
HIP103682 & 5973 & 39& 56& 4.27 & 0.01 & 0.47 & 0.06 & 0.61 & --0.12 & 0.05 \\
HIP103692 & 5708 & 77& 77& 4.21 & 0.02 & 0.29 & 0.09 & 1.26 & 0.04 & 0.08 \\
HIP104672 & 6114 & 37& 62& 4.11 & 0.02 & --0.07 & 0.10 & 1.62 & 0.06 & 0.05 \\
\enddata
\tablecomments{A solar Fe abundance of A(Fe)$_{\odot}$ = 7.5 was adopted from \citet{asplund09}, and A(P)$_{\odot}$ = 5.43 from this work (see \ref{subsection::solarp}).}
\tablecomments{The microturbulence ($\xi$) uncertainties are discussed in \ref{subsec::minesweeper_input}.}
\tablecomments{(This table is available in its entirety in machine-readable form.)}
\end{deluxetable*}

The stars selected from APOGEE DR16 have greater differences between our results and the parameters from \citet{jonsson20}. However, these systematic differences are similar in size when APOGEE results are compared to other abundance surveys \citep{jonsson18}. The largest difference is in surface gravity, however the P I lines are relatively unaffected to changes on log(g) compared to changes in $\teff$ and the uncertainty on the synthetic spectrum fits (e.g. uncertainties in \citealt{maas19} and values listed in Table \ref{table::abund_uncer}).

We also compare our derived [Fe/H] ratios to previous studies in Fig. \ref{fig::feh_comp} and statistics in Table \ref{table::atmo_param_comparisons}. We do not test our [Fe/H] ratios with the PASTEL catalog to avoid averaging [Fe/H] abundance measurements from different methodologies and complicating interpretations of potential systematic uncertainties with our method. We use Fe I abundances from \citet{bensby14} to match our methodology, and re-normalize all literature results to the \citet{asplund09} solar scale. We find no significant systematic offsets and uncertainties similar to our quoted measurement uncertainties. Since we specifically compare other \citet{bensby14} abundances to P in section \ref{subsec::p_evol_other_elements}, we perform a linear regression on the [Fe/H] differences vs. [Fe/H] from this work (left panel of Fig. \ref{fig::feh_comp}) to look for systematic uncertainties between methodologies. We find a slope of 0.00 $\pm$ 0.02, with a p-value = 0.89, indicating there is no statistically significant slope when comparing the [Fe/H] difference vs. [Fe/H]. This suggests our [Fe/H] derivation methodology is consistent with \citet{bensby14} for both the metal rich and metal poor stars in our sample.

\begin{deluxetable*}{ c c c c c c c c}
\tablewidth{0pt} 
\tabletypesize{\footnotesize}
\tablecaption{Atmospheric Parameter Comparisons \label{table::atmo_param_comparisons}} 
 \tablehead{\colhead{Lit. Source} & \colhead{Mean $\Delta$ $\teff$} &  \colhead{$\sigma$ $\Delta$ $\teff$} & \colhead{Mean $\Delta$log(g)} &  \colhead{$\sigma$ $\Delta$ log(g)} & \colhead{Mean $\Delta$ [Fe/H]} &  \colhead{$\sigma$ $\Delta$[Fe/H]} & \colhead{Ref.} \\ \colhead{} & \colhead{(K)} & \colhead{(K)} & \colhead{(dex)}& \colhead{(dex)} & \colhead{(dex)}& \colhead{(dex)} & \colhead{}}
\startdata
Bensby+14  &  --31 & 99 & -0.06 & 0.12 & --0.03 & 0.10  &(1) \\
Casgrande+11 & --28 & 84 & -0.04 & 0.11 & 0.02 & 0.11  &(2) \\
PASTEL Catalog & --1 & 84 & --0.01 & 0.07 & \nodata & \nodata&  (3) \\
APOGEE DR16 & --58 & 58 & --0.16 & 0.14 & 0.05 & 0.08 & (4) \\
\hline
\multicolumn{8}{c}{Stars With Precise 2MASS Photometry} \\
\hline
Bensby+14  &  --21 & 92 & --0.04 & 0.10 & --0.02 & 0.09 & (1)\\
Casgrande+11 & --27 & 78 & -0.03 & 0.11 & 0.02 & 0.10 & (2)\\
PASTEL Catalog & 2 & 76 & 0 & 0.07 & \nodata & \nodata & (3)\\
APOGEE DR16 & --30 & 42 & --0.16 & 0.17 & 0.06 & 0.08 & (4) \\
\enddata
\tablecomments{Sources (1) \citealt{bensby14}; (2) \citealt{casagrande11}; (3) \citealt{soubiran16}; (4) \citealt{jonsson20}. The stars with precise IR photometry only include stars with two or more good JHK photometries (errors $<$ 0.2 mag). The listed differences are literature values subtracted from this work (e.g., $\mathrm{[Fe/H]} _{\mathrm{This Work}}$ - $\mathrm{[Fe/H]} _{\mathrm{Literature}}$) }
\end{deluxetable*}

\subsection{Solar P Abundance}
\label{subsection::solarp}
We obtained a twilight spectrum from the HPF on the HET, to compare our Fe (discussed in section \ref{subsec::minesweeper_input}) and P derivation methodology to the solar values. While the telluric lines are strengthened in the twilight spectrum, the P I line profile was not contaminated and the line profile was symmetric.  We measured the solar phosphorus abundances using the P I line and calculated the uncertainty on the synthetic spectrum fit. We found a A(P)$_{\odot}$ = 5.43 $\pm$ 0.02, which is slightly less than the value of 5.46 from \citet{caffau07}, greater than the measurement of 5.41 from \citet{asplund09}, but matches the solar system values of A(P) = 5.43 dex \citep{lodders21}. We therefore adopt a solar phosphorus abundance of A(P)$_{\odot}$ = 5.43 dex.

Previous studies have found systematic differences between twilight solar spectra to reflected light spectra, specifically a 0$\%$ - 4$\%$ decrease in line depth depending on the angular distance from the sun \citep{gray00}, and a 1$\%$ - 2$\%$ decrease with UVES data \citep{molaro08}. The equivalent widths for affected lines are also thought to decrease by similar percentages \citep{gray00}.  A 1$\%$ or 4$\%$ increase in line strength would lead to an abundance change of 0.004 dex or 0.02 dex respectively. The effects on the solar abundance from use of a twilight spectrum are likely minimal in the specific case of the weak P I line compared to the uncertainty on the fit, however other solar abundances were not derived using the twilight spectrum.

\subsection{Phosphorus Abundance Methodology}
\label{subsection::phos_abun_method}
Phosphorus abundances were measured by fitting synthetic spectra to the 10529.52 $\ang$ feature. The wavelength, excitation potential ($\chi$), and log(gf) values for the P I absorption line were adopted from \citet{berzinsh97} and are listed in Table \ref{table::linelist}. Synthetic spectra were generated using MOOG \citep{sneden73} and MARCS atmospheric models \citep{gustafsson08}. Each spectrum was shifted to the rest frame before abundance analysis using radial velocity and barycentric corrections derived from template matching within \texttt{iSpec} \citep{blanco-cauresma14,blanco-cauresma19}. Some stars rotate rapidly enough that additional smoothing was required. We adopted projected rotation velocities from the Geneva Copenhagen survey \citep{holmberg07} for stars with vsini $\geq$ 5 km s$^{-1}$. Stars with smaller reported rotation speeds did not have their abundances change significantly when the v sin i was included in the fit. A grid of synthetic spectra of the P I line were created in steps of 0.01 dex, over a range of -1 $<$ [P/Fe] (dex) $<$ 1.5. The best fitting model was chosen using $\chi ^{2}$ minimization and examples of model spectra are shown for three stars in Fig. \ref{fig:ir_spec}. The final, best fitting phosphorus abundances for each star are listed in Table \ref{table::params_abuns}. 

The uncertainty on the model fit is estimated with a Monte Carlo method; the observed fluxes are varied using Gaussian random numbers with standard deviations equal to the SNR of the spectrum at 10600 $\ang$, following the methodology of \citet{maas19}. The model then is re-fit, the process repeated for 10,000 iterations, and the abundances at the 16$\%$ and 84$\%$ percentiles of the distribution are taken as the 1 $\sigma$ uncertainties. 

We also estimated the impact from the uncertainties on the atmospheric parameters on the fitted abundance. We created separate models at the $\pm$1 sigma level for each of the atmospheric parameters. For example, if a star had an $\teff$ = 6000K $\pm$ 100K, we would create models at 5900 and 6100K, determine the best fitting synthetic spectrum in a newly created grid, then use the differences to the [P/Fe] ratio at 6000K to determine the uncertainties. We assume the uncertainties are independent of each parameter and we add the final values in quadrature. The most dominant sources of uncertainty is from the model fit, as shown in Table \ref{table::abund_uncer}.

\begin{deluxetable}{ c c c}
\tablewidth{0pt} 
\tabletypesize{\footnotesize}
\tablecaption{Average Phosphorus Abundance Uncertainties \label{table::abund_uncer}} 
 \tablehead{\colhead{} & \colhead{$< \delta$[P/Fe]$>$}  & \colhead{$\sigma$ ($\delta$[P/Fe])} \\ \colhead{} & \colhead{(dex)}  & \colhead{(dex)} }
\startdata
$\teff$ Upper &  0.03 & 0.01 \\
$\teff$ Lower   & 0.03 & 0.01 \\
log(g) &  0.03 & 0.01 \\
$\mathrm{[Fe/H]}$ &  0.01 & 0.01 \\
$\xi$ & 0.01 & 0.03 \\
Fit Upper &  0.04 & 0.01 \\
Fit Lower &  0.04 & 0.01 \\
\enddata
\tablecomments{Average abundance uncertainties due to errors in various parameters were calculated independently and represented as $< \delta$[P/Fe]$>$.}
\end{deluxetable}

\begin{figure}[tp!]
	\centering 
 	\includegraphics[trim=0cm 0cm 0cm 0cm, scale=.45]{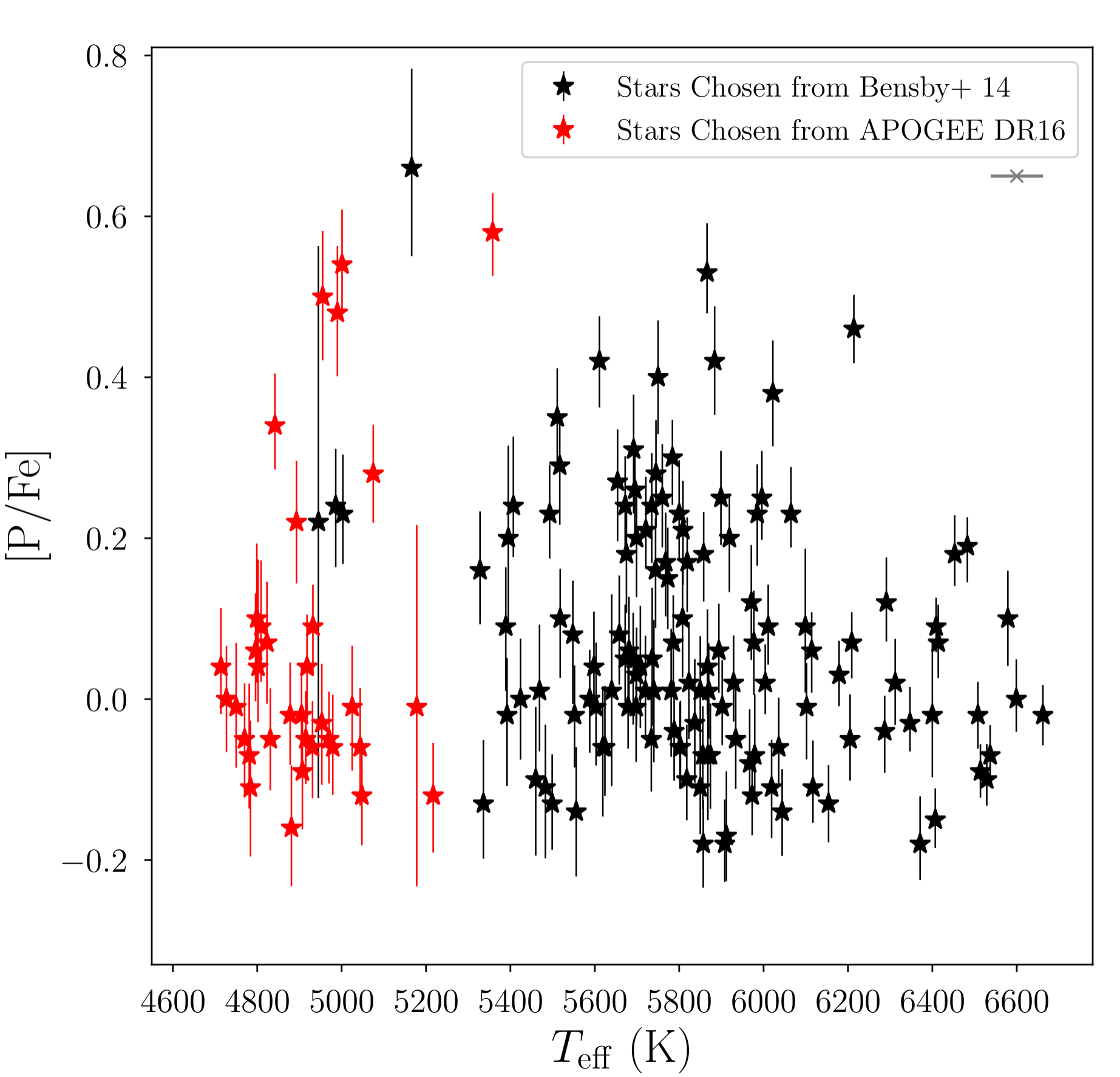}
	\caption{Comparison of [P/Fe] to $\teff$ from this work. Black stars are targets selected from \citet{bensby14} and red stars are those chosen from APOGEE DR16. The gray `x' in the upper right corner is a representative error bar for $\teff$.  \label{fig::p_fe_teff} }
	\end{figure}

We also examined how [P/Fe] ratios vary as a function of the stellar $\teff$. A relationship between $\teff$ and our abundance measurements can reveal unknown systematic effects from our abundance measurement methodology, such as deviations from LTE, or from difficulties with the linelist (e.g. unidentified blends). We find no significant change in [P/Fe] ratios across our $\teff$ range, as shown in Fig. \ref{fig::p_fe_teff}. Other studies of similar P I features in the Y-band have found little difference in [P/Fe] ratio for Hyades dwarfs and giants, consistent with our results \citep{maas19}. Additionally, consistent [P/Fe] ratios for stars spanning 4700 K - 6700 K suggests no detectable, $\teff$ dependent unidentified blend is likely affecting the 10529.52 $\ang$ line.

\begin{figure}[tp!]
	\centering 
 	\includegraphics[trim=0cm 0cm 0cm 0cm, scale=.47]{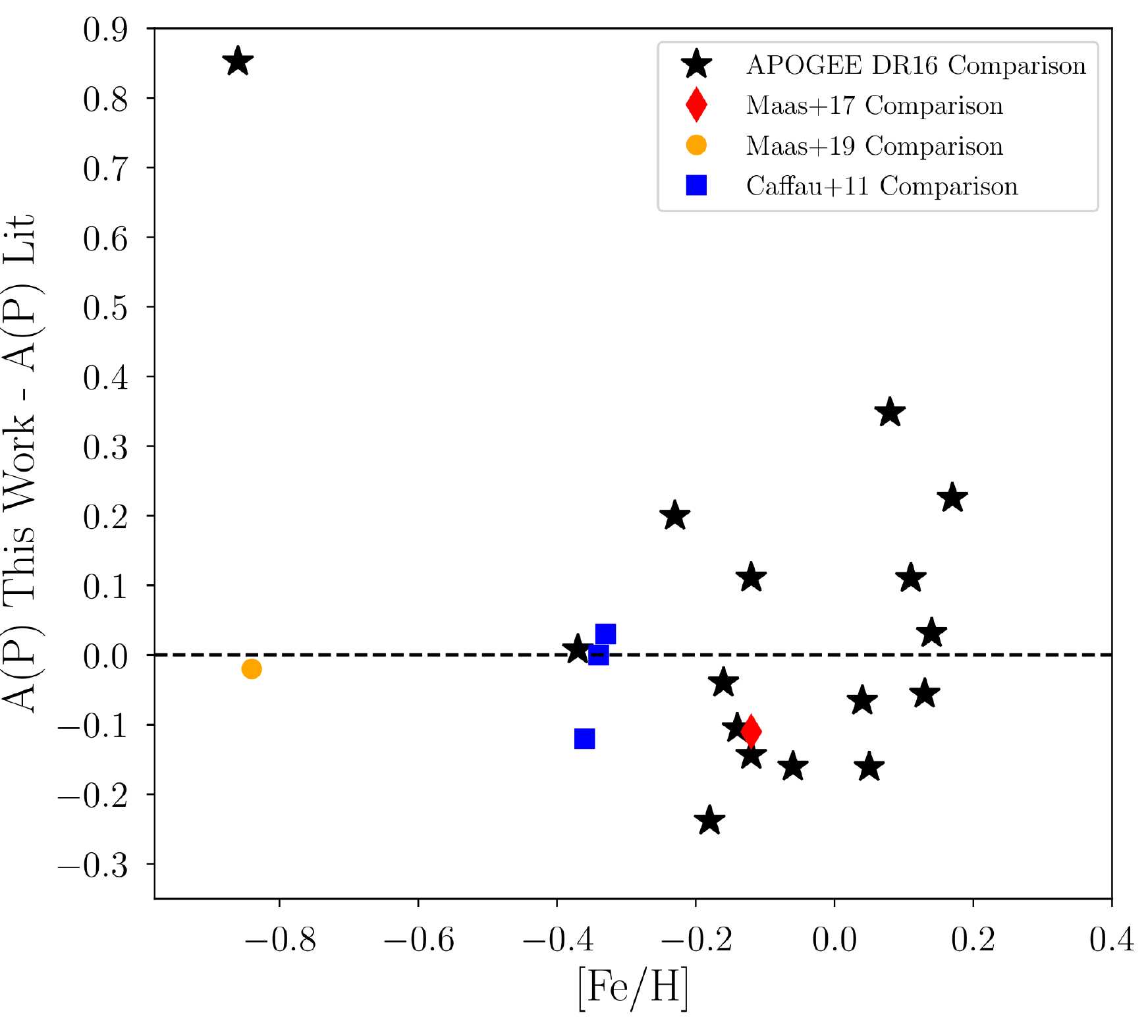}
	\caption{A(P) abundances were compared from this study to abundances from APOGEE DR 16 (black stars), from \citet[][red diamonds]{maas17}, from \citet[][orange circle]{maas19}, and from \citep[blue squares]{caffau11}. \label{fig::p_fe_comp} }
	\end{figure}

We compare our P abundances to measurements from the literature, including 12 stars with reported [P/Fe] ratios from APOGEE DR16 \citep{jonsson20}, and plot the differences in Fig. \ref{fig::p_fe_comp}. There are six stars in common with other Y-band P I studies: three with \citet[][HIP 10306, HIP 23941, HIP 107975]{caffau11}, two in common with \citet[][HIP 174147, HIP 38625]{maas19}, and one from \citet[][HIP 174160]{maas17}. We compare our abundances from Table \ref{table::params_abuns} with the values from those studies in Fig. \ref{fig::p_fe_comp}. We do not include a comparison with HIP 38625 since the [P/Fe] uncertainty from \citet{maas19} is 0.22 dex. We removed the solar normalization and ratio to [Fe/H] to ensure only a comparison of the phosphorus abundances were examined between the various studies. In the case of APOGEE DR16, we adopt an A(P)$_{\odot}$ = 5.36 from \citet{grevesse07} then apply a correction of 0.183 dex, from Table 4 of \citet{jonsson20} to match solar metallicity stars in the solar neighborhood. We find an average difference of --0.04 $\pm$ 0.06 dex for the five P I values and for APOGEE we find a difference of 0.0 $\pm$ 0.17 dex, excluding the one outlier star, HD 81762. A larger sample is needed for to conduct a more detailed comparison with APOGEE.

\section{Discussion}
\label{sec::discussion}

\subsection{Comparing Thin Disk to Thick Disk Stars}

\label{subsec::thin_thick_disk}

Abundances differences between the thin and thick disk stars exist for the $\alpha$-elements, indicative of the different chemical histories of each population of stars in the Milky Way (e.g. \citealt{edvardsson93,bensby14,hayden15}). We examined whether phosphorus behaved similarly to the $\alpha$-elements in the thin and thick disk for our stars, as suggested by \citet{maas19}. To do this, we first derived U, V, W velocities for each of our stars using Gaia eDR3 positions, parallaxes, proper motions, and radial velocities \citep{gaia21} with the \texttt{pyia} code \citep{price-whelan_2018} in order to properly account for the correlated uncertainties on the final galactocentric velocity components. 

We generated 1000 sample parameters for each star using the Gaia covariance matrices, computed the resulting positions and velocities in the galactocentric frame, and adopted the standard deviation and mean values as the best sets of velocities and associated uncertainties using \texttt{pyia} \citep{price-whelan_2018}. The coordinate transformation into the galactocentric frame were calculated using \texttt{astropy} python package \citep{astropy13}. Finally, the velocities were corrected to the local rest frame by removing a rotation velocity of 238 km s$^{-1}$ \citep{Marchetti18}.

\begin{figure}[tp!]
	\centering 
 	\includegraphics[trim=0cm 0cm 0cm 0cm, scale=.48]{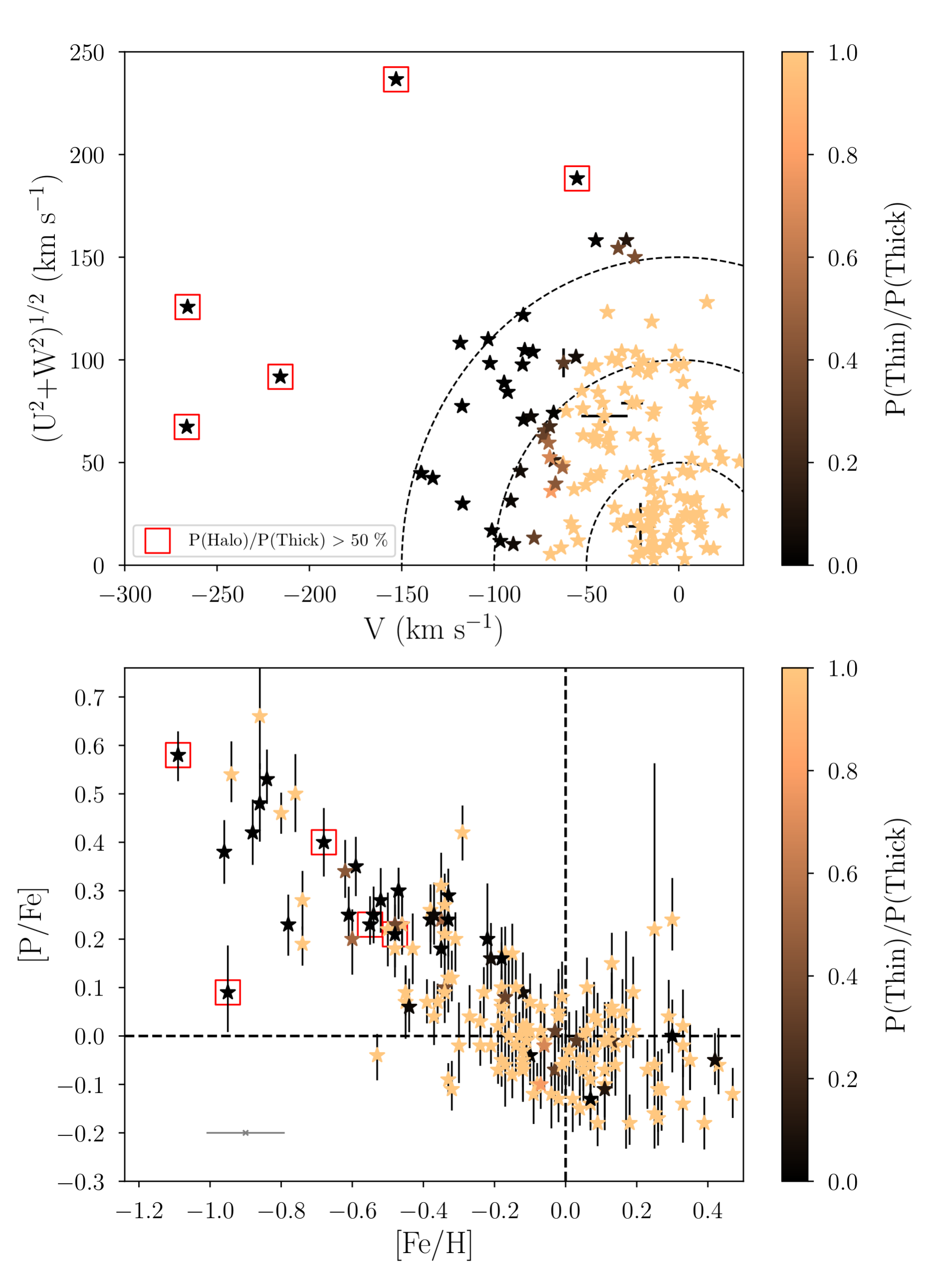}
	\caption{Top Panel: The galactocentric velocity V and the sum of the squares of the U and W velocity components are plotted for the observed stars. Dashed lines represent radii at 50 km s$^{-1}$, 100 km s$^{-1}$, and 150 km s$^{-1}$. Bottom Panel: [P/Fe] ratio compared to the [Fe/H] ratio. In both panels, the color bar stars the thin disk to thick disk probability ratio with black stars being likely thick disk members and gold stars are probable thin disk stars. The gray `x' in the bottom left corner is a representative [Fe/H] error bars. Stars in red boxes have halo-to-thick disk probabilities above 50$\%$.  \label{fig:toomre} }
	\end{figure}

The galactocentric velocities were then used to determine the likely stellar population assignment for each sample star using the methodology of \citet{ramirez13}. The use of galactocentric velocities to identify thin disk, thick disk, and halo stars is consistent with other methods in the literature (e.g. \citealt{bensby14}) and has given similar results to abundance studies that have derived population probabilities using orbital parameters \citep{franchini20}. The UVW velocities, thin disk, thick disk, and halo membership probabilities are listed in Table \ref{table::kinematics}. A Toomre diagram and the [P/Fe] vs [Fe/H] relation with thin/thick disk color-coding are shown in Fig. \ref{fig:toomre}. 

We made selection cuts to quantify any difference between the two stellar populations. Following \citet{bensby14}, we selected stars with P(Thin)/P(Thick) $>$ 10 for thin disk stars and P(Thin)/P(Thick) $<$ 0.1 for thick disk stars to ensure stars with inconclusive population probabilities did not affect the analysis. We examined the metallicity range between --0.9 $<$ [Fe/H] $<$ --0.1 where the [$\alpha$/Fe] ratio begins to diverge between thin and thick disk stars, but before [$\alpha$/Fe] plateaus at low [Fe/H] ratios. We also recalculated UVW velocities given in Table 4 of \citet{maas19} with the methodology above and added those stars to the sample to increase the number of stars used in the analysis. We made a final cut on [Fe/H] uncertainty and removed stars with high [Fe/H] uncertainties ($\sigma$[Fe/H] $>$ 0.1 dex).  

\begin{figure}[tp!]
	\centering 
 	\includegraphics[trim=0cm 0cm 0cm 0cm, scale=.5]{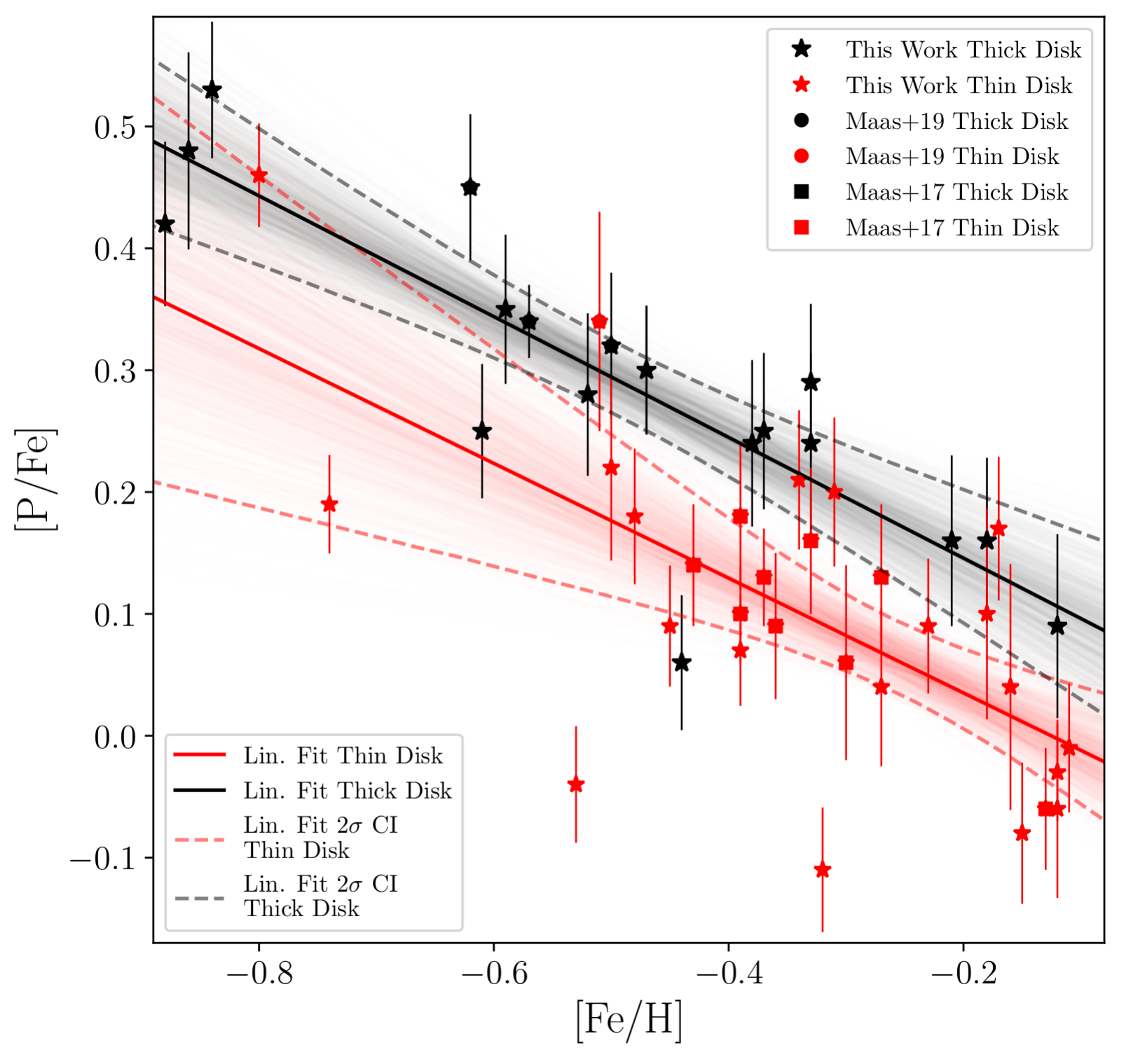}
	\caption{[P/Fe] vs. [Fe/H] from multiple studies. Stars that are highly likely thin disk members (red points) and thick disk members (black points) are plotted (as defined in section \ref{subsec::thin_thick_disk}). Red and black lines are linear fits (lin. fits) to the thin disk population and thick disk population respectively, dashed lines represent the 2$\sigma$ confidence intervals, and the fainter lines represent 1,000 samples from the MCMC fit. Data from this work are represented as stars, data from \citet{maas17} are squares, and data from \citet{maas19} are circles.} \label{fig:thin_thick}
	\end{figure}

The resulting sample of stars are shown in Fig. \ref{fig:thin_thick}. Initial inspection reveals a small difference between thin and thick disk stars although there is scatter in the data. We fit a linear line to the thin and thick disk abundance ratios to trace the evolution of phosphorus in each population. The best fitting model was calculated using a Markov Chain Monte Carlo simulation performed using the \texttt{emcee} package \citep{foremanmackey13}. 

We used 10,000 steps with 50 walkers and the fits are shown as dashed lines in Fig. \ref{fig:thin_thick}. The resulting slope and intercept parameters for thin disk and thick disk stars are given in Table \ref{table::lin_fit_params}. The reported uncertainties are the 16$\%$ and 84$\%$ percentiles on the posterior distribution of each quantity. We re-compute the linear fit using only the stars from this work and find the thin disk slope flattens and all quantities are more uncertain, but a [P/Fe] difference between both stellar populations is still found. 

\begin{deluxetable*}{ c c c c c c c}
\tablewidth{0pt} 
\tabletypesize{\footnotesize}
\tablecaption{Thin Disk and Thick Disk Linear Fit Parameters \label{table::lin_fit_params}} 
 \tablehead{\colhead{Sample}  & \colhead{Slope} & \colhead{$\sigma$ Slope High}  & \colhead{$\sigma$ Slope Low} & \colhead{Intercept} & \colhead{$\sigma$ Intercept High}  & \colhead{$\sigma$ Intercept Low} \\ \colhead{} &  \colhead{} & \colhead{}  & \colhead{} & \colhead{(dex)} & \colhead{(dex)}  & \colhead{(dex)} }
\startdata
Thin Disk: All Stars &  --0.47 & 0.11 & 0.12 & --0.06 & 0.03 & 0.03 \\
Thick Disk: All Stars   &  --0.50 & 0.14 & 0.07 & 0.05 & 0.05 & 0.05 \\
Thin Disk: This Work Only &   --0.37 & 0.15 & 0.17 & --0.05 & 0.03 & 0.04 \\
Thick Disk: This Work Only  &  --0.46 & 0.24 & 0.08 & 0.03 & 0.05 & 0.05 \\
\enddata
\tablecomments{All stars include those of \citet{maas17,maas19} that meet the criteria described in section \ref{subsec::thin_thick_disk}.}
\end{deluxetable*}

We have also found that the [P/Fe] ratio decreases with increasing [Fe/H] at similar rates for both populations. We compute the 2$\sigma$ confidence intervals (CIs) using the combined abundances from this work and from \citet{maas17,maas19}. The CIs are plotted in Fig. \ref{fig:thin_thick} and we find that that the difference between both fits exceeds the 2$\sigma$ CIs between --0.57 $<$ [Fe/H] $<$ --0.13, although it is not significant at the 3$\sigma$ level. Beyond this metallicity range we lack enough measurements to confirm any offset from thin disk and thick disk stars. From our linear fitting parameters, we measure a [P/Fe] offset of 0.11 dex between both populations. The chemical evolution model of \citet{kobayashi11} predicts an offset between thin and thick disk stars of approximately 0.1 dex, similar to the observed value. The difference in [P/Fe] ratio between the thin and thick disk is similar to the behavior of other $\alpha$-elements and suggest P is produced in CCSNe.

\begin{deluxetable*}{ c c c c c c c c c c}
\tablewidth{0pt} 
\tabletypesize{\footnotesize}
\tablecaption{Stellar Kinematics and Galactic Population Membership Probabilities\label{table::kinematics}} 
 \tablehead{\colhead{Star Name} & \colhead{U} &  \colhead{$\sigma$ U} & \colhead{V} &  \colhead{$\sigma$ V} & \colhead{W} &  \colhead{$\sigma$ W} &  \colhead{P(Thin Disk)} &  \colhead{P(Thick Disk)}  & \colhead{P(Halo)}  \\ \colhead{} & \colhead{(km s$^{-1}$)} & \colhead{(km s$^{-1}$)} & \colhead{(km s$^{-1}$)} & \colhead{(km s$^{-1}$)}& \colhead{(km s$^{-1}$)}& \colhead{(km s$^{-1}$)}& \colhead{}& \colhead{}& \colhead{}}
\startdata
HIP101346	&	--40.6	&	0.2	&	--14.6	&	0.2	&	--0.5	&	0.1	&	0.99	&	0.01	&	0.00 	\\
HIP102610 &	-10.3	&	0.1	&	-15.0	&	0.1	&	18.8	&	0.1	&	0.99	&	0.01	&	0.00	\\
HIP102838 &	-96.2	&	0.1	&	-45.0	&	0.1	&	14.2	&	0.1	&	0.69	&	0.30	&	0.00	\\
HIP103682 &	5.8	&	0.1	&	-16.8	&	0.1	&	4.6	&	0.1	&	0.99	&	0.01	&	0.00	\\
HIP103692 &	5.8	&	0.1	&	-16.8	&	0.1	&	4.6	&	0.1	&	0.99	&	0.01	&	0.00	\\
HIP104672 &	-61.3	&	0.1	&	-44.8	&	0.1	&	-10.9	&	0.1	&	0.88	&	0.12	&	0.00	\\
\enddata
\tablecomments{UVW values are galactocentric velocities corrected to the Local Standard of Rest.}
\tablecomments{(This table is available in its entirety in machine-readable form.)}
\end{deluxetable*}

\subsection{The Galactic Evolution of Phosphorus}
\label{subsec::chem_evol}
\begin{figure}[tp!]
	\centering 
 	\includegraphics[trim=0cm 0cm 0cm 0cm, scale=.4]{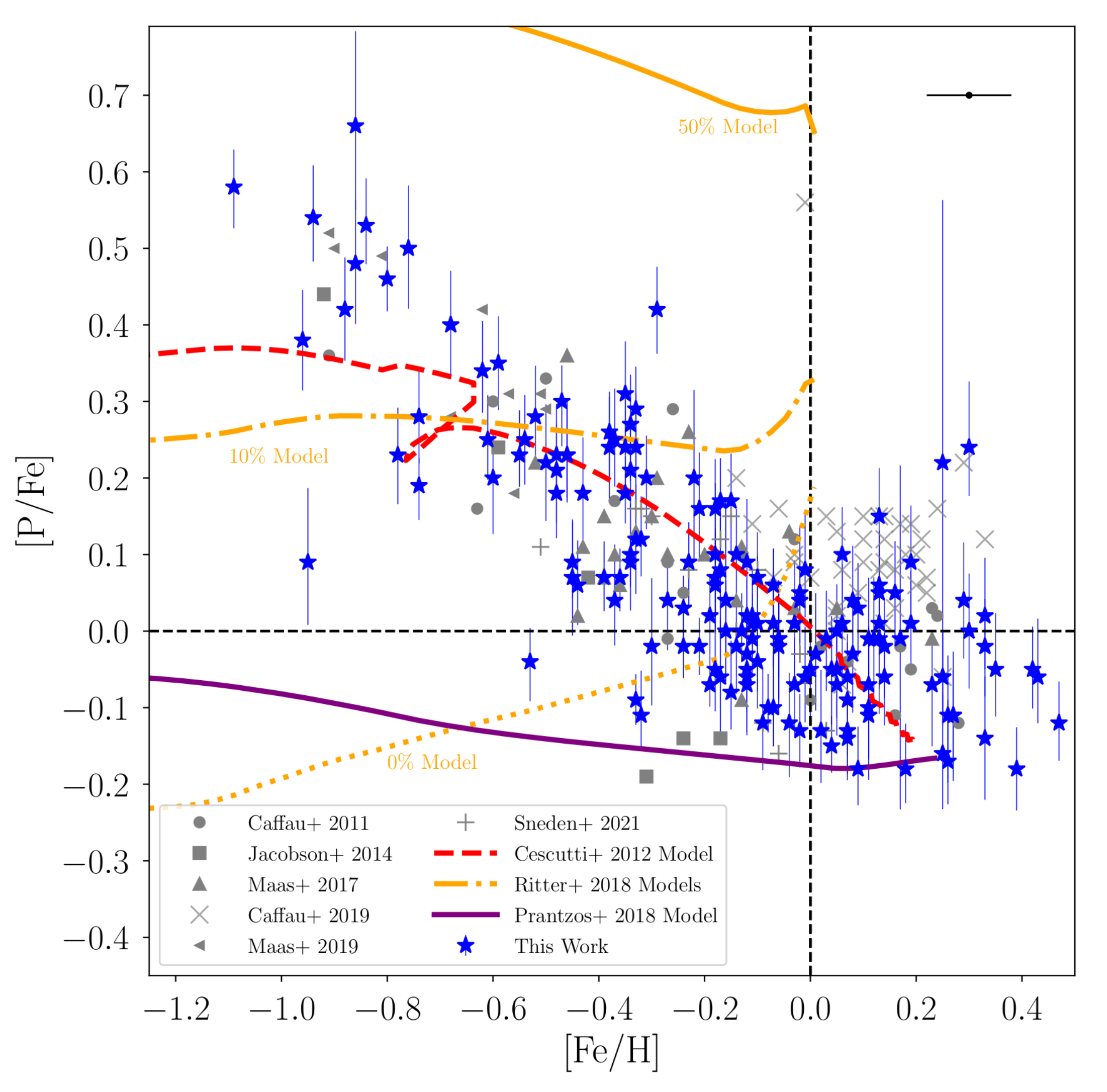}
	\caption{[P/Fe] vs [Fe/H] are plotted for this work (blue stars) and from other studies as gray markers. We include phosphorus abundance ratios as circles from \citet{caffau11}, as squares from \citet{jacobson14}, as triangles from \citet{maas17}, as rotated triangles from \citet{maas19}, as `x's from \citet{caffau19}, and plusses from \citet{sneden21}. The solar abundance ratios are shown as black dashed lines and [Fe/H] representative error bars for this work in the upper right of the plot. The chemical evolution model from \citet{cescutti12} is a red dashed line and the model from \citet{prantzos18} is a purple solid line. Three \citet{ritter18} models with different amount of stars undergoing C-O shell mergers are included as orange lines, with the 0$\%$ model as a dotted line, the 10$\%$ model as a dot-dashed line, and a 50$\%$ model as a solid line.  \label{fig::p_fe_chem_evol} }
	\end{figure}

Our phosphorus abundance ratio measurements are shown in Fig. \ref{fig::p_fe_chem_evol} and compared with chemical evolution models from \citet{cescutti12,prantzos18,ritter18} to further test if P behaves similarly as the $\alpha$-elements. Each model has different assumptions on phosphorus yields and have been tuned to match stars in the solar neighborhood. The best fitting model is from \citet{cescutti12}, which includes massive star yields from \citet{kobayashi06}, hypernovae, and a metallicity-independent P yield enhancement. However, additional abundance measurements in stars with metallicities between --1.2 $<$ [P/Fe] $<$ --0.5 show the model under-predicts the [P/Fe] ratio at lower [Fe/H]. A more complicated solution than a metallicity-independent yield increase is therefore needed to explain the production of phosphorus in more metal-poor stars. 

The \citet{prantzos18} model includes updated yields of massive rotating stars in addition to CCSN and Type Ia yields. While the addition of these yields have greatly improved models with other odd-Z elements (e.g. K), the \citet{prantzos18} model under-produces [P/Fe] and does not match the slope of the data. The C shell ingestion in convective O shells may also enhance the yields of the odd-Z elements in massive stars. \citet{ritter18} investigated the how potential yield enhancements would affect chemical evolution models if 0$\%$, 10$\%$,  50$\%$, and 100$\%$ of stars in the Galaxy had yields enhanced from the O-C shell interactions. Three of those models are plotted in Fig. \ref{fig::p_fe_chem_evol} and we find no model perfectly matches the slope of the [P/Fe] ratios as a function of metallicity (similar to \citealt{maas19}). However, the 10$\%$ model approximately matches from --0.6 $<$ [Fe/H] $<$--0.4 and more enhanced yields are needed at lower metallicities. 

In addition to chemical evolution models, we compare our abundances to different P abundance studies from multiple different surveys. We find our abundances are in general agreement with studies that used the IR Y-band and H-band P I features \citep{caffau11,maas17,maas19,sneden21}. The abundances of \citet{caffau19} of stars near solar metallicity are slightly offset from our abundance results however. This survey uses Y-band IR features as well and the discrepancy may be due to differences in atmospheric parameter choice or measurement methodology. Finally, we note that some high [Fe/H] stars that utilize UV P I features are lower than our abundance ratios, however the UV lines saturate and are highly uncertain near [Fe/H] $\sim$ --0.2 \citep{roederer14,jacobson14}.

\subsection{The Evolution of Phosphorus Compared to Other Elements}	
\label{subsec::p_evol_other_elements}
\begin{figure*}[tp!]
	\centering 
 	\includegraphics[trim=0cm 0cm 0cm 0cm, scale=.37]{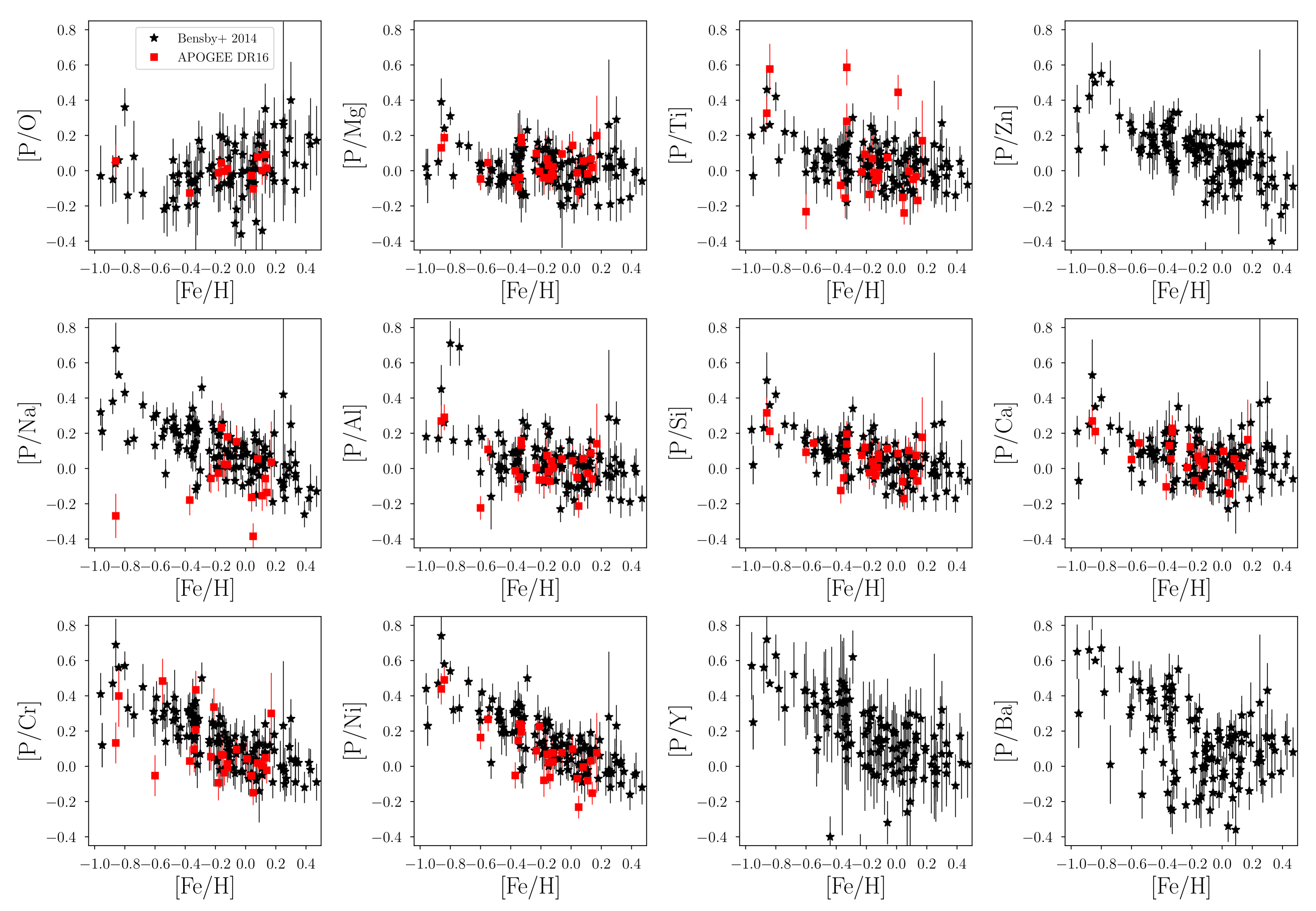}
	\caption{[P/X] vs [Fe/H] ratios are plotted using P abundances from this work, and other abundances from \citet{bensby14} (as black circles) and APOGEE DR16 (red squares). \label{fig::p_x_analysis} }
	\end{figure*}

We compared phosphorus abundances to $\alpha$-elements, iron-peak elements, and s-process elements from \citet{bensby14,jonsson20} to explore possible sites of P production. We chose to use abundances of other studies due to the lack of O, Al, and s-process element lines in the Y-band (e.g. \citealt{sneden21}). Different [P/X] ratio are plotted as a function of [Fe/H] in Fig. \ref{fig::p_x_analysis}.

\textit{$\alpha$-Element Comparison:} We compared phosphorus abundance measurements to O, Mg, Si, and Ca. We find that [P/O] increases with increasing metallicity, [P/Mg] is approximately constant from --1.1 $<$ [P/Mg] $<$ 0.5, and [P/Si] (and [P/Ca] to an extent) are slightly decreasing with increasing [Fe/H]. Both Si and Ca are thought to be made in explosive burning in CCSN and have significant Type Ia contributions. The constancy with Mg and slight decreasing slope with Ca and Si suggests P is mostly made in CCSN as expected from nucleosynthesis predictions.

\textit{Iron-Peak Element Comparison:} We compared phosphorus to both Ni, Cr, and Zn (in addition to Fe in Fig. \ref{fig::p_fe_chem_evol}). We find strongly decreasing [P/Cr] and [P/Ni], suggesting that P is made relatively quickly by massive stars and not in type Ia supernova and further resembles the evolution of the $\alpha$-elements.

\textit{Odd-Z Element Comparison:} We compared phosphorus to Na and Al and found [P/Na] ratios increasing more significantly with [Fe/H] relative to [P/Al] ratios, where [P/Al] evolves more similarly to the $\alpha$-elements. Na is thought to have strongly metallicity dependent yields  \citep{andrews17} in massive stars and phosphorus appears to have a different behavior over our [Fe/H] range. 

\textit{S-Process Element Comparison:} Finally, phosphorus abundance ratios with Y and Ba are explored. While there is significant scatter in the data due to larger uncertainties on the s-process element abundances from \citet{bensby14}, we still find [P/Y] and [P/Ba] ratios decrease with increasing metallicity. The [P/Y] and [P/Ba] ratios suggest that phosphorus does not correlate with disk star abundances of elements produced by the s-process in AGB stars. 

We find that the evolution of P most closely resembles the $\alpha$-elements, especially Mg, over our observed [Fe/H] range. The more significant the contributions of delayed processes, such as Type Ia SN and AGB stars production, the strong the [P/X] ratio decreases as [Fe/H] increases. Our abundance ratio comparisons are in agreement with our previous analysis that P is likely mostly produced in CCSNe.

\subsection{Stars with Anomalous [P/Fe] Ratios}

The [P/Fe] ratio for some of the stars in our sample are multiple standard deviations away from the average phosphorus abundance for a given [Fe/H] ratio. We highlight three of these stars and discuss possible origins.

\subsubsection{HIP 100792}

HIP 100792 star has been identified as a low-$\alpha$ member of the Galactic halo suggesting the star is a member of an accreted system \citep{fuhrmann98,nissen10,hawkins15}. \citet{nissen10,nissen11} have also measured [Na/Fe] =  --0.12 and an [Mn/Fe] = --0.25 for this star and these low Na and Mn abundances are characteristic of the accreted halo and especially stars in the Gaia-Sausage-Enceladus system \citep[or GSE,][]{belokurov18,haywood18,helmi18,das20}. We also examine the kinematics of this star, specifically the vertical angular momentum component (L$_{z}$) and radial action (J$_{r}$), using the package \texttt{galpy} \citep{bovy15}.

We found L$_{z}$ = --360.7 kpc km s$^{-1}$ and J$_{r}$ = 567.4 kpc km s$^{-1}$ (or  $\sqrt{\mathrm{J}_{r}}$ = 23.8  [kpc km s$^{-1}$]$^{\frac{1}{2}}$  ) using the potential \texttt{MWPotential2014} and the Staeckel approximation (described in \citealt{binney12}). The kinematics agree with GSE kinematics from other studies (e.g. J$_{r}$ $>$ 500 kpc km s$^{-1}$ and L$_{z}$ $<$ 500 kpc km s$^{-1}$ from \citealt{yuan20}) and is nearly consistent with the selection criteria used in other GSE analysis such as 30 $\leq$ $\sqrt{\mathrm{J}_{r}}$  $\leq$ 50 (kpc km s$^{-1}$)$^{\frac{1}{2}}$
and --500 $\leq$ L$_{z}$  $\leq$ 500 kpc km s$^{-1}$ \citep{feuillet21,buder22}. We note the dynamical selection of GSE stars alone does not lead to a pure sample of GSE members and can include up to $\sim$ 20$\%$ non-accreted contaminants \citep{bonifacio21}. HIP 100792 therefore is chemically and dynamically similar to GSE stars, suggesting this star is likely a member of the same accreted population. 

HIP 100792 has a [Fe/H] = -0.95 $\pm$ 0.09 with a [P/Fe] = 0.09 $\pm$ 0.1.  This star is the only identified member of the low-$\alpha$ halo in our sample and the [P/Fe] ratio is $\sim$ 0.3 - 0.5 dex offset from similar thin and thick disk stars. The low P abundance for this one star is intriguing and motivates a larger P survey in accreted halo stars. 

\subsubsection{HIP 18833}

HIP 18833 has an [Fe/H] = --0.53 $\pm$ 0.05 with an [P/Fe] = --0.04 $\pm$ 0.05. While the phosphorus abundance on the star is low, other chemical abundances do not suggested this is an accreted star. The [X/Fe] ratios for Mn, Al, and Na all appear characteristic of disk stars: [Mn/Fe] = --0.02 \citep{feltzing07}, [Na/Fe] = 0.07, and [Al/Fe]=0.20 \citep{bensby14} -- although [Al/Fe] has a large uncertainty of $\pm$ 0.18. The thin/thick disk probability ratio is also 63.6, indicating this star is very likely a thin disk member based on kinematics. Finally, \citet{bensby14} finds a [Mg/Fe] = 0.03 $\pm$ 0.11 for this star, similar to the [P/Fe] = --0.07 $\pm$ 0.05. The low abundance ratios suggests the low P abundance is likely related to the stars chemical history as opposed to an errant measurement.

\subsubsection{HIP 95447}

HIP 95447 has an [Fe/H] = 0.3 $\pm$ 0.07 with an [P/Fe] = 0.24$\pm$ 0.08. The $\alpha$-element abundance ratios for this star are not enhanced, for example [Mg/Fe] = 0.02 $\pm$ 0.1 \citep{bensby14}. However, there is a large spread of $T_{\mathrm{eff}}$ for this star in the literature. The average $\teff$ in the PASTEL catalog is 5600 $\pm$ 147 K \citep{soubiran16} and we found and $\teff$ of 5407$^{+69}_{-34}$ K. HIP 95447 demonstrates some photometric and chromospheric variability as well \citep{lockwood07}. Broad-band photometry may not reflect the star at time of observation with HPF. A lower $\teff$ would require higher abundances to fit the P I line potentially creating the offset. The most likely cause of outlier [P/Fe] ratio for this star is an unknown systematic effect on the atmospheric parameters. 

\subsection{Molar Phosphorus Ratios}

P is a critical element for multiple biological reactions. To better connect astronomical studies of P abundances to other disciplines, \citet{hinkel20} analyzed stellar abundances as molar ratios. The authors then specifically compared ratios of key elements for life of C, N, Si, and P and found the Sun has distinctive abundances compared to nearby stars (lower P/C molar ratios) and that the Earth is relatively enriched with phosphorus. The low refractory-to-volatile element ratios are consistent with other studies that compared the solar composition to solar twins \citep{bedell18}. We extend that analysis and have examined P relative to other key elements for life by computing the molar ratios of P/C and N/C for our stars.

\begin{figure}[tp!]
	\centering 
 	\includegraphics[trim=0cm 0cm 0cm 0cm, scale=.43]{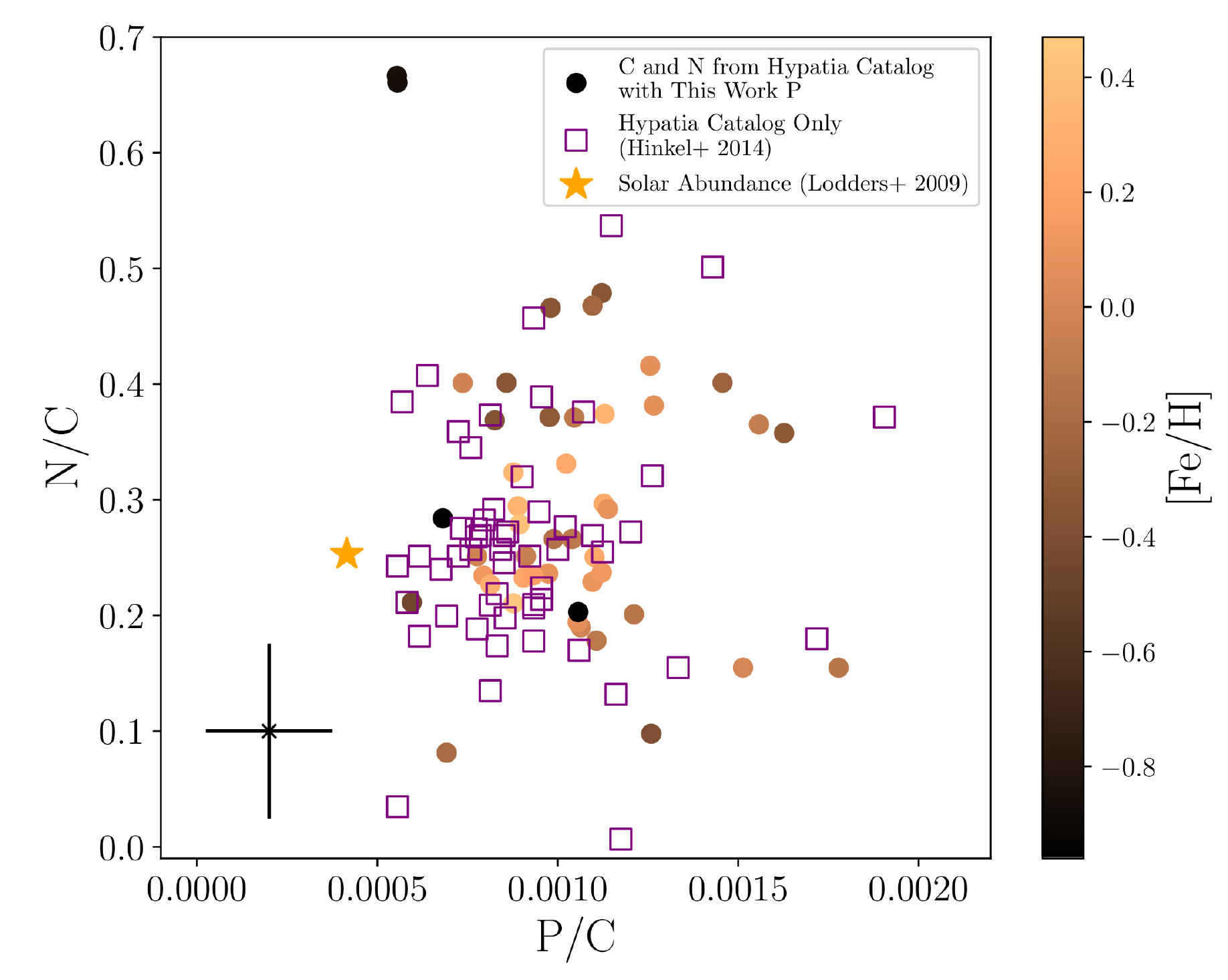}
	\caption{Molar N/C ratios compared to P/C ratios. The orange star is the solar abundance from \citet{lodders09}. All stars in the plot have C and N abundances adopted from the Hypatia Catalog \citep{hinkel14}. Stars with P abundances measured in this work are shown as circles and a color bar denotes the [Fe/H]. Literature P abundances from Hypatia are plotted as open-faced purple squares. A representative error bar is also shown in the bottom left corner.  \label{fig::p_c_molar} }
	\end{figure}

We obtained C and N abundance ratios from the Hypatia Catalog \citep{hinkel14} for selected stars in our sample. We chose stars that had log(g) $>$ 3.5 dex to avoid giants that may have surface abundances affected by the first dredge-up (additional N or removed $^{12}$C).  In total, 46 dwarfs in our sample had both literature N and C abundances listed in the Hypatia Catalog. We used the mean abundance for stars with multiple C or N measurements and we adopted the standard deviation from multiple abundance measurements as the abundance uncertainty. The abundances are adopted directly from the Hypatia Catalog and not corrected for any difference in atmospheric parameters. Molar ratios were calculated using Eq. 2 from \citet{hinkel20} and are plotted in Fig. \ref{fig::p_c_molar} for our sample. We have also added the stars with previously measured C, N and P abundances (P abundances from \citealt{caffau11,caffau16,maas17,maas19,caffau19}) from the Hypatia Catalog, and previously analyzed in \citet{hinkel20}, to compare with our sample.

Our results are largely consistent with the molar ratios shown in \citet{hinkel20}. We also calculated the solar N/C and P/C ratios using abundances from \citet{lodders09} and we find the P/C ratio of the sun is offset from other field stars; likely due to a slight C enrichment relative to other stars in the solar neighborhood \citep{bedell18}. The 46 stars with molar ratios include four stars with [Fe/H] $<$ --0.8; all other stars are between --0.45 $<$ [Fe/H] $<$ 0.47. The four relatively metal-poor stars include the two outliers at N/C $\sim$ 0.6, HIP 38625 and HIP 17147. The two other metal-poor stars are more consistent with the rest of the sample and include the halo star HIP 100792 and the thick disk star HIP 104659 which have N/C values of 0.28 and 0.20 respectively. 

A combination of intrinsic abundance scatter, systematic uncertainties in atmospheric parameter derivation between different abundance studies, uncertainties in the molar ratios, and chemical evolution of more metal poor stars may be creating the large observed scatter in the molar ratios. Additional P abundances are critical to understand the P/C evolution with [Fe/H] and to understand the range of important elements for life in nearby stars. A significant variation of C/P and N/C ratios may impact life as P may be strongly partitioned in planetary cores \citep{stewart07}. Continuing to characterize the variations of elements such as P, is the first step to constraining what bio-essential elements may be available in other exoplanets \citep{hinkel20}. 

\section{Conclusions}
\label{sec::conclusions}
While P is an important element for life and is potentially made in multiple nucleosynthesis sites, few stellar abundances exist to test theories of P production. We measured P abundances in 163 stars using observations obtained with HPF on the HET. Our sample was selected from the previous abundance studies of \citet{bensby14} and APOGEE DR16 \citep{jonsson20}. High SNR observations were obtained for stars that ranged over --1.09 $<$ [Fe/H] $<$ 0.47. Atmospheric parameters were derived using \mines\ and a combination of optical and infrared photometry, Gaia parallaxes, and Fe I equivalent widths. We derived a P abundances by fitting a grid of synthetic spectra, created using MOOG \citep{sneden73} with MARCS model atmospheres \citep{gustafsson08}, to the P I feature at 10529.52 $\mbox{\AA}$. We summarize the main conclusions of our analysis below:

\begin{enumerate}
\item{We demonstrated the P I line at 10529.52 $\ang$ gave consistent abundances for both red giants and FGK dwarf stars. We find that our P abundances are generally consistent with previous works, however there was significant scatter when comparing 12 stars in common with APOGEE DR16, consistent with P being relatively uncertain in APOGEE \citep{jonsson20}}.
    
\item{This sample of P abundances supports the results of \citet{maas19} that chemical evolution models do not fit the observed abundances of stars in the solar neighborhood. The best fitting model from \citet{cescutti12} uses P massive star yields that have been enhanced by a factor of 2.75. However, the model does not fit the most metal-poor stars in our sample. Yield enhancements from C-O shell ingestion in massive stars may be a promising way to resolve discrepancy between models and abundances \citep{ritter18}.}

\item{We identify members of the Milky Way thin and thick disk using UVW velocities and membership criteria from \citet{ramirez13}. We find a [P/Fe] difference of $\sim$ 0.1 dex between thin and thick disk stars, providing additional evidence that phosphorus is primarily made in CCSNe.}
    
\item{Comparisons of phosphorus to multiple difference elements in Fig. \ref{fig::p_x_analysis}. We find phosphorus evolves similarly to Mg and [P/X] ratios with iron-peak or s-process elements decrease with increasing [Fe/H]. The behavior of [P/Mg] compared to [P/Y] and [P/Ni,Cr] suggests P is primarily made in CCSNe and not in time-delayed processes for disk stars.}

\item{We computed molar P/C and N/C fractions for a sub-sample of 46 dwarf stars. We found that the solar P/C ratio is offset relative to field stars, agreeing with measurements of \citet{hinkel14,hinkel20}.We also find a large scatter in the observations that may be due to different star-to-star intrinsic abundances and differences in atmospheric parameter measurement methodology in the dwarf star sample.} 

\item{We find one star, HIP 100792, is relatively deficient in P compared to stars at similar metallicities. HIP 100792 is also a likely halo star and member of the GSE system based on kinematics and chemistry. Further observations of low-$\alpha$ halo stars are needed to confirm that HIP 100792 is not an anomaly.}        

\item{Our analysis of 163 stars overall suggests P is primarily produced in CCSNe and has evolved in lockstep with other $\alpha$-elements.}

\end{enumerate}

\software{MOOG (v2019; \citealt{sneden73}), \texttt{scipy} \citep{scipy20}, \texttt{numpy} \citep{walt11}, \texttt{matplotlib} \citep{hunter07}, \texttt{astropy} \citep{astropy13}, \texttt{galpy} \citep{bovy15}, \texttt{emcee} \citep{foremanmackey13}, \texttt{pyia} \citep{price-whelan_2018}, \texttt{dustmaps} \citep{green19}, \texttt{MINESweeper} \citep{cargile20}, \texttt{iSpec} \citep{blanco-cauresma14,blanco-cauresma19}, \texttt{ThePayne} \citep{ting19}, Goldilocks HPF Pipeline \url{https://github.com/grzeimann/Goldilocks_Documentation}}

\section{Acknowledgements}

The research shown here acknowledges use of the Hypatia Catalog Database, an online compilation of stellar abundance data as described in Hinkel et al. (2014, AJ, 148, 54), which was supported by NASA’s Nexus for Exoplanet System Science (NExSS) research coordination network and the Vanderbilt Initiative in Data-Intensive Astrophysics (VIDA). This research has made use of the NASA Astrophysics Data System Bibliographic Services and this research has made use of the SIMBAD database, operated at CDS, Strasbourg, France. This publication makes use of data products from the Two Micron All Sky Survey, which is a joint project of the University of Massachusetts and the Infrared Processing and Analysis Center/California Institute of Technology, funded by the National Aeronautics and Space Administration and the National Science Foundation. This publication makes use of data products from the Wide-field Infrared Survey Explorer, which is a joint project of the University of California, Los Angeles, and the Jet Propulsion Laboratory/California Institute of Technology, funded by the National Aeronautics and Space Administration. This work has made use of data from the European Space Agency (ESA) mission {\it Gaia} (\url{https://www.cosmos.esa.int/gaia}), processed by the {\it Gaia} Data Processing and Analysis Consortium (DPAC, \url{https://www.cosmos.esa.int/web/gaia/dpac/consortium}). Funding for the DPAC has been provided by national institutions, in particular the institutions participating in the {\it Gaia} Multilateral Agreement. The Hobby-Eberly Telescope (HET) is a joint project of the University of Texas at Austin, the Pennsylvania State University, Ludwig-Maximilians-Universität München, and Georg-August-Universität Göttingen. The HET is named in honor of its principal benefactors, William P. Hobby and Robert E. Eberly. These results are based on observations obtained with the Habitable-zone Planet Finder Spectrograph on the HET. The HPF team was supported by NSF grants AST-1006676, AST-1126413, AST-1310885, AST-1517592, AST-1310875, AST-1910954, AST-1907622, AST-1909506, ATI 2009889, ATI-2009982, and the NASA Astrobiology Institute (NNA09DA76A) in the pursuit of precision radial velocities in the NIR. The HPF team was also supported by the Heising-Simons Foundation via grant 2017-0494. We thank the referee Dr. P. Bonifacio for
suggested improvements to the manuscript. Z. G. M and N. R. H. are partially supported by a NASA ROSES-2020 Exoplanet Research Program Grant (20-XRP20 2-0125). KH acknowledges support from the National Science Foundation grant AST-1907417 and AST-2108736. KH is partly support from the Wootton Center for Astrophysical Plasma Properties funded under the United States Department of Energy collaborative agreement DE-NA0003843. This work was performed in part at the Simons Foundation Flatiron Institute's Center for Computational Astrophysics during KH's tenure as an IDEA Fellow.

\FloatBarrier

\bibliographystyle{aasjournal}
\bibliography{phos_references}

\end{document}